\documentclass[titlepage]{amsart}
\pdfoutput=1

\usepackage[foot]{amsaddr}
\usepackage{lmodern}
\usepackage{fix-cm}
\usepackage{cite}
\usepackage{amsmath,amssymb,amsfonts}
\usepackage{mathtools}
\usepackage[dvipdfm]{graphicx}
\usepackage{bmpsize} 
\usepackage{textcomp}
\usepackage{xcolor}
\usepackage{latexsym}
\usepackage[colorlinks=true,linkcolor=blue]{hyperref}
\usepackage{array}
\usepackage{subcaption}

\def\Hone{$\mathcal{H}$}
\def\Htwo{$\mathcal{H}^2$}
\newcommand{\pp}[2]{{{}_#2}{#1}}
\newcommand{\comp}[1]{\overline{#1}} 

\def\x{\mathbf{x}}
\def\y{\mathbf{y}}

\usepackage{algorithm}
\usepackage[noend]{algpseudocode}
\renewcommand\algorithmicdo{}        
\algrenewcommand\alglinenumber[1]{\footnotesize #1}
\algrenewcommand\algorithmicindent{0.9em}%
\algblockdefx{ForAllp}{EndForAllp}[1] {\textbf{in parallel for} #1}

\makeatletter
\ifthenelse{\equal{\ALG@noend}{t}}%
  {\algtext*{EndForAllp}}
  {}%
\makeatother

\algnewcommand\algorithmicforeach{\textbf{for each}}
\algdef{S}[FOR]{ForEach}[1]{\algorithmicforeach\ #1\ \algorithmicdo}
\algnewcommand{\LineComment}[1]{\State \(\triangleright\) #1}

\newcommand{\batch}[1]{{#1_{\scriptscriptstyle \vert\kern-0.24ex\vert\kern-0.24ex\vert}}}
\newcommand{\cal}[1]{\mathcal{#1}}

\begin{document}

\title[H2Opus]{H2Opus: A distributed-memory multi-GPU software package for non-local operators}

\author[S.Zampini et. al.]{Stefano Zampini}
\address{Extreme Computing Research Center, Computer, Electrical and Mathematical Science and Engineering Division, King Abdullah University of Science and Technology, Thuwal, Saudi Arabia}
\email{stefano.zampini@kaust.edu.sa}
\author[]{Wajih Boukaram}
\email{wajih.boukaram@kaust.edu.sa}
\author[]{George Turkiyyah}
\email{george.turkiyyah@kaust.edu.sa}
\author[]{Omar Knio}
\email{omar.knio@kaust.edu.sa}
\author[]{David E. Keyes}
\email{david.keyes@kaust.edu.sa}

\begin{abstract}
Hierarchical $\mathcal{H}^2$-matrices are asymptotically optimal representations for the discretizations of non-local operators such as those arising in integral equations or from kernel functions. Their $O(N)$ complexity in both memory and operator application makes them particularly suited for large-scale problems. As a result, there is a need for software that provides support for distributed operations on these matrices to allow large-scale problems to be represented. In this paper, we present high-performance, distributed-memory GPU-accelerated algorithms and implementations for matrix-vector multiplication and matrix recompression of hierarchical matrices in the $\mathcal{H}^2$ format.

The algorithms are a new module of H2Opus, a performance-oriented package that supports a broad variety of $\mathcal{H}^2$ matrix operations on CPUs and GPUs. Performance in the distributed GPU setting is achieved by marshaling the tree data of the hierarchical matrix representation to allow batched kernels to be executed on the individual GPUs. MPI is used for inter-process communication. We optimize the communication data volume and hide much of the communication cost with local compute phases of the algorithms. Results show near-ideal scalability up to 1024 NVIDIA V100 GPUs on Summit, with performance exceeding 2.3 Tflop/s/GPU for the matrix-vector multiplication, and 670 Gflops/s/GPU for matrix compression, which involves batched QR and SVD operations.

We illustrate the flexibility and efficiency of the library by solving a 2D variable diffusivity integral fractional diffusion problem with an algebraic multigrid-preconditioned Krylov solver and demonstrate scalability up to 16M degrees of freedom problems on 64 GPUs.
  
\end{abstract}

\maketitle

\date{\today}


\section{Introduction}
\label{sec:intro}

Many of the large dense matrices that appear in scientific computing, machine learning, integral equations, and other applications are in fact \emph{data sparse}. They may be approximated, to arbitrary accuracy, with a memory footprint that is much smaller that the prohibitive $O(N^2)$ cost of the dense representation. Data sparsity is a consequence of the fact that certain blocks of the matrix may be represented by factorizations whose ranks are smaller than the block dimensions. These low rank blocks of the matrix may be of different sizes and may appear in general position in the matrix.  The resulting representations are known as hierarchical matrices.

Hierarchical matrices are often used to compress kernel matrices, where a general point set and an explicit kernel function are given. This is a natural setting for N-body problems, boundary and volume integral methods, spatial statistics, regression and discretizations of fractional and non-local operators among others. Hierarchical matrices are also effective in settings where explicit kernels do not exist. Inverses of discretization of differential operators, Schur complements, and Hessians of PDE-constrained optimization problems also result in matrices that have a structure with many low rank blocks. General linear algebraic operations are possible in the compressed representations, presenting opportunities for tackling challenging computations and provide scalable solutions for operations that would be otherwise computationally infeasible with the dense format.

Hierarchical matrices come in different variants. A basic version uses a fixed blocking of the matrix (so-called weak admissibility partitioning), where every block touches the diagonal and is stored as a separate low rank factorization. The asymptotic memory footprint of this representation is $O(k N \log N)$, where $k$ is a representative rank of the blocks. 
Unfortunately, this basic representation requires substantial memory in practice due to both the $\log$ factor and the large ranks $k$ generally needed to reach reasonable target accuracy, as a result of the rigid partitioning of the matrix. 
This can be improved on in two ways. The first eliminates the logarithmic factor in the complexity estimates by expressing the block factorizations in a nested basis. 
The second improvement, which allows a substantial reduction in the effective $k$, allows both dense and low rank blocks of different granularity to appear in general position in the matrix, essentially allowing the matrix structure itself to be adapted to a given problem (strong admissibility partitioning).  The hierarchical matrix variant with these two improvements is referred to as the \Htwo{} format, and reaches the optimal asymptotic memory footprint $O(k N)$. Given that memory is often the primary constraining resource in high performance computing environments, the representation of choice for large scale problems is a general, nested basis, strong admissibility, \Htwo{} representation. \Htwo{} matrices were originally motivated primarily by the needs of integral equations \cite{hackbusch02}, but their algebraic nature has allowed their application in a variety of other contexts, including, for example, the representation of Dirichlet-to-Neumann operators \cite{gillman14,gillman15}, particulate flows in periodic geometries of arbitrary shapes \cite{marple16}, Hessians of PDE-constrained problems \cite{boukaram20b}.

There is tremendous interest in the development of high-performance software libraries for constructing and operating on hierarchical matrices, driven naturally by their compelling small memory footprint and lower asymptotic operational complexity. Much like quality libraries such as FMM3D \cite{fmm3d} have made it possible to spread the use of FMM technology to broad classes of problems, similar libraries for \Htwo{} matrices---which are algebraic generalizations of FMM---would also allow a broader use of this technology for tackling dense matrix problems that might otherwise be prohibitively expensive.  Software packages that are not performance-oriented but more oriented towards readability and conciseness include \cite{ho20,massei20,ambikasaran19,h2lib}.  A shared-memory high-performance package for \Htwo{} matrices is presented in \cite{hua21}. A high quality software for distributed-memory environments that targets large scale problems is STRUMPACK \cite{strumpack}. 
Distributed memory algorithms for constructing and operating on these matrices were proposed in \cite{ida14,borm08,yamazaki19}.
Dynamic run-time systems to manage the scheduling of the operations of the irregular hierarchical matrix structure in a more convenient fashion through explicit task graphs are presented in \cite{aliaga17, borm19}. 
Manycore algorithms were presented in \cite{ohshima18,zaspel19}, and BiCG solvers on GPU clusters were demonstrated in \cite{yamazaki18}. 
Multicore and manycore methods for operating on hierarchical matrices including matrix-vector multiplication are discussed in \cite{biros16,biros17,erlandson20}. 
Algorithms for high performance computing environments suitable for large scale problems are presented in \cite{rouet16,ghysels17,yu18,rebrova18,yu19}. 


\begin{figure}[ht]
  \begin{center}
    \includegraphics[width=0.6\textwidth]{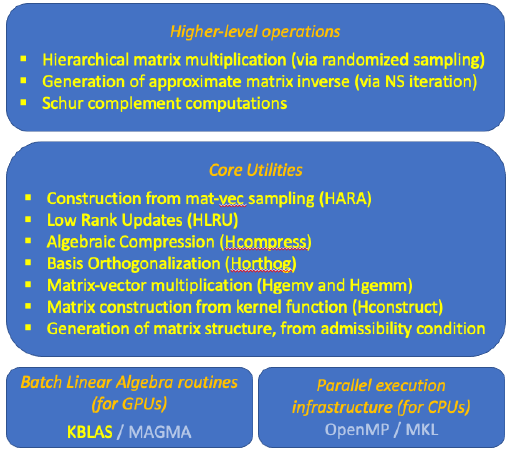}
    \caption{Functionality of the H2Opus library}
    \label{fig:arch}
  \end{center}
\end{figure}

H2Opus \cite{h2opus} is an open source, GPU-accelerated, package for hierarchical matrix computations in the \Htwo{} format. It supports a number of core operations (see Figure \ref{fig:arch}) for constructing these matrices from a kernel and admissibility conditions, performing BLAS-like operations 
including matrix-vector multiplication, basis orthogonalization, and recompression. It also provides facilities for adding a (globally) low rank matrix to an \Htwo{} matrix, as well as the ability to construct an \Htwo{} matrix from randomized sampling operations, essentially generalizing the construction of globally low rank matrix decomposition \cite{halko11} to the hierarchically low rank case. These core operations form the building blocks for higher level operations that perform matrix-matrix multiplication, generate approximate inverses iteratively, and compute Schur complements in various applications \cite{boukaram19b,boukaram20b}. The package runs on CPUs (with explicit optimizations for Intel and AMD CPUs) as well as on GPUs with the portability enabled by a lower level layer that either uses batch BLAS operations specialized to run directly on the GPUs when manycore hardware is available, or alternatively uses OpenMP to obtain batch parallelism in multicore environments. In addition, the library also supports running on the NEC-SX Vector Engine. H2Opus has interfaces to the PETSc \cite{petsc-user-ref,petsc-web-page} package, allowing the use of the extensive facilities of the latter for manipulating distributed-memory (sparse) matrices arising from the discretization of PDEs. Efficient solvers are also available in the so called Tile Low Rank format \cite{boukaram21}.

In this paper, we extend the H2Opus package with distributed-memory multi-GPU operations for two of its core capabilities:  matrix-vector multiplication (and the related multiplication of a hierarchical matrix by multiple vectors), and matrix recompression. Single-GPU versions of these algorithms were presented in \cite{boukaram19a}. 
Matrix-vector multiplication is a key building block of Krylov-type iterative solvers, appearing in the inner loops of nonlinear and eigenvalue solvers. Therefore improving its performance can substantially reduce overall time to solution. The multi-vector case is also a performance critical kernel in many contexts such as randomized hierarchical matrix-matrix multiplication, and block-Krylov methods, such as those exposed by PETSc \cite{jolivet2021ksphpddm}. The core matrix-vector multiplication can benefit from processing multiple vectors concurrently; the additional arithmetic intensity made available, relative to looping over the bandwidth-limited operation of single vector multiplication, allows substantially higher absolute performance to be achieved.  
Matrix recompression is also another key building block for \Htwo{} matrix operations, closely resembling the truncation operations on dense matrices and matrix blocks. Recompression is needed when an initial \Htwo{} matrix approximation is generated using a polynomial interpolation or other non-optimal bases, and algebraic recompression step is relied on to produce a storage-optimal representation for the desired accuracy. It is also needed when BLAS3 operations are performed on \Htwo{} matrices. These operations generally increase the apparent rank of various blocks, which would then need to be recompressed to maintain the optimal complexity.

We present high performance and scalable $O(N)$ implementations of these algorithms that demonstrate near-ideal scalability  on up to 1024 NVIDIA V100 GPUs, with performance exceeding 2.3 Tflop/s per GPU. At this scale, a dense kernel matrix of size 536M$\times$536M, represented as a hierarchical matrix to an accuracy of $\epsilon=10^{-7}$ can be multiplied by a single vector in $17$ ms and by 64 vectors concurrently in $66$ ms, achieving 85\% of the peak efficiency of batched GEMM operations.  This pushes the state of the art beyond the results of \cite{yu18,yu19,rouet16}. The compression operation also achieves near-optimal scalability with number of GPUs, with matrices of size 67M$\times$67M compressed by a factor of 6 in their low rank memory (from an accuracy of $10^{-6}$ to an accuracy of $10^{-3}$) in around $320$ ms on 64 GPUs.


The algorithms are supported by distributed data structures for representing, accessing, and operating on hierarchical matrices with nested bases. These data structures have elements similar to those found in parallel multigrid representations, with analogous restriction and prolongation transfer operators, since the low rank blocks and their bases appear at different levels of granularity and are naturally stored at multiple levels of a hierarchy. The data structures also have patterns similar to those of parallel sparse matrix representations because at every level of the hierarchy, the low rank blocks appear in general locations forming essentially a block sparse matrix whose block sizes are on the order of the block ranks, $k$.  The matrix structure has a bounded sparsity constant~\cite{grasedyck03,borm10} which is exploited to optimize overall inter-process communication volume and to hide much of the MPI communication cost with concurrent local compute phases of the algorithms. Single node performance is obtained by marshaling the tree data structures in a way to allow optimized batched dense linear algebra kernels to be invoked. 

We finally demonstrate the use of the algorithms in the scalable solution of a 2D variable-coefficient integral fractional diffusion equation. We use the distributed-memory algorithms to construct and compress the dense operator to an accuracy of $10^{-6}$. The solution uses a Krylov method with the matrix-vector multiplication done by our distributed algorithm and the
preconditioning done by a classical (non-fractional) diffusion equation which is solved by a distributed-memory algebraic multigrid method in PETSc. The results show that all aspects of the solver, including the setup of the hierarchical representation of the dense operators, the computation of the volume ``Dirichlet'' conditions, the setup of the preconditioner, and the work per iteration, scale linearly with problem size.  The solver also exhibits a dimension-independent number of iterations and results in near-ideal weak scalability up to grids of size $4096\times4096$ on 64 GPUs.

The rest of this paper is organized as follows. Section \ref{sec:hrep} describes the distributed representation of the data structures of an \Htwo{} matrix. Section \ref{sec:dist_hgemv} describes the basic distributed matrix-vector multiplication operations and Section \ref{sec:dist_hgemv_opt} presents the key optimizations to minimize inter-process communication volume and hide latencies. Section \ref{sec:compression} describes the recompression algorithms which shares many of the algorithmic structures of the matrix-vector multiplication algorithms but relies on batch QR and SVD instead of batch GEMM in lower level linear algebra. Section \ref{sec:results} presents the distributed multi-GPU scalability results for the matrix-vector multiplication and recompression algorithms as well as the integral fractional diffusion solver. We conclude with future potential directions in Section \ref{sec:conclusions}.

\section{Distributed Data Structures for \Htwo{} Matrices}
\label{sec:hrep}

\subsection{\Htwo{}  Matrix Structure}

The representation of an \Htwo{} matrix exploits two kinds of hierarchies: a hierarchy of blocks at different levels of granularity and a hierarchy of bases that is used to represent the data in the blocks. At every level of the hierarchy the low rank blocks essentially form a block sparse matrix. In addition, there is a block sparse matrix that stores the dense blocks of the matrix that are not amenable to low rank representations. 

\begin{figure}[!ht]
\begin{subfigure}{\columnwidth}
  \centering
  \hspace*{-8pt} \includegraphics[width=0.8\textwidth,clip,trim={4.3cm 16.7cm 22.2cm 0}]{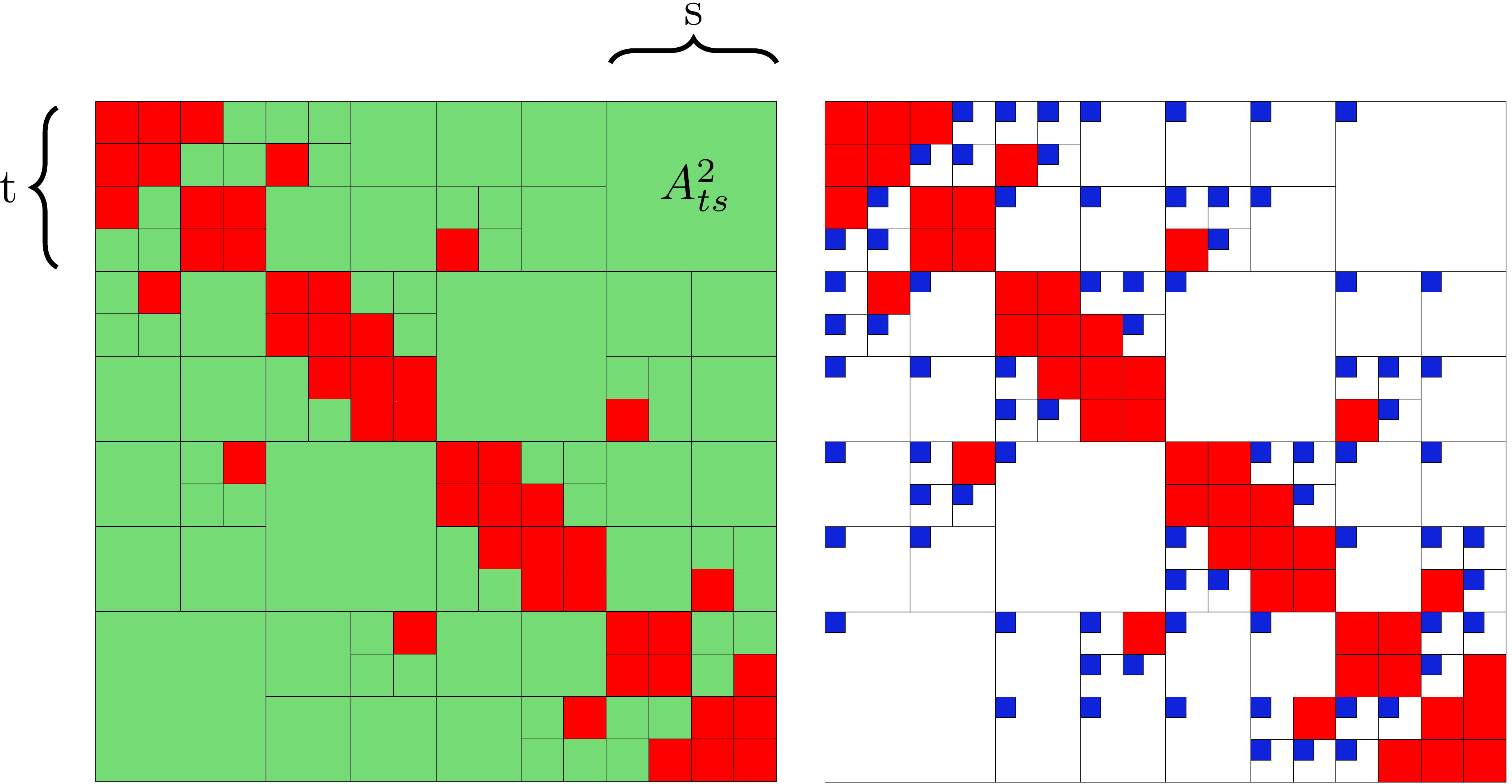}
  \vspace*{-4pt}
  \caption{Leaves of the matrix tree: (left) in the \Hone{} format, leaf blocks such as the one labeled at level 2 of the tree, $A^2_{ts}$, are stored as low rank factorizations; (right) in the \Htwo{} format, low rank blanks are represented by much smaller coupling blocks $S^2_{ts}$.}
  \label{fig:hmatrix_leaves}
\end{subfigure}

\begin{subfigure}{\columnwidth}
  \centering
  \includegraphics[height=2cm]{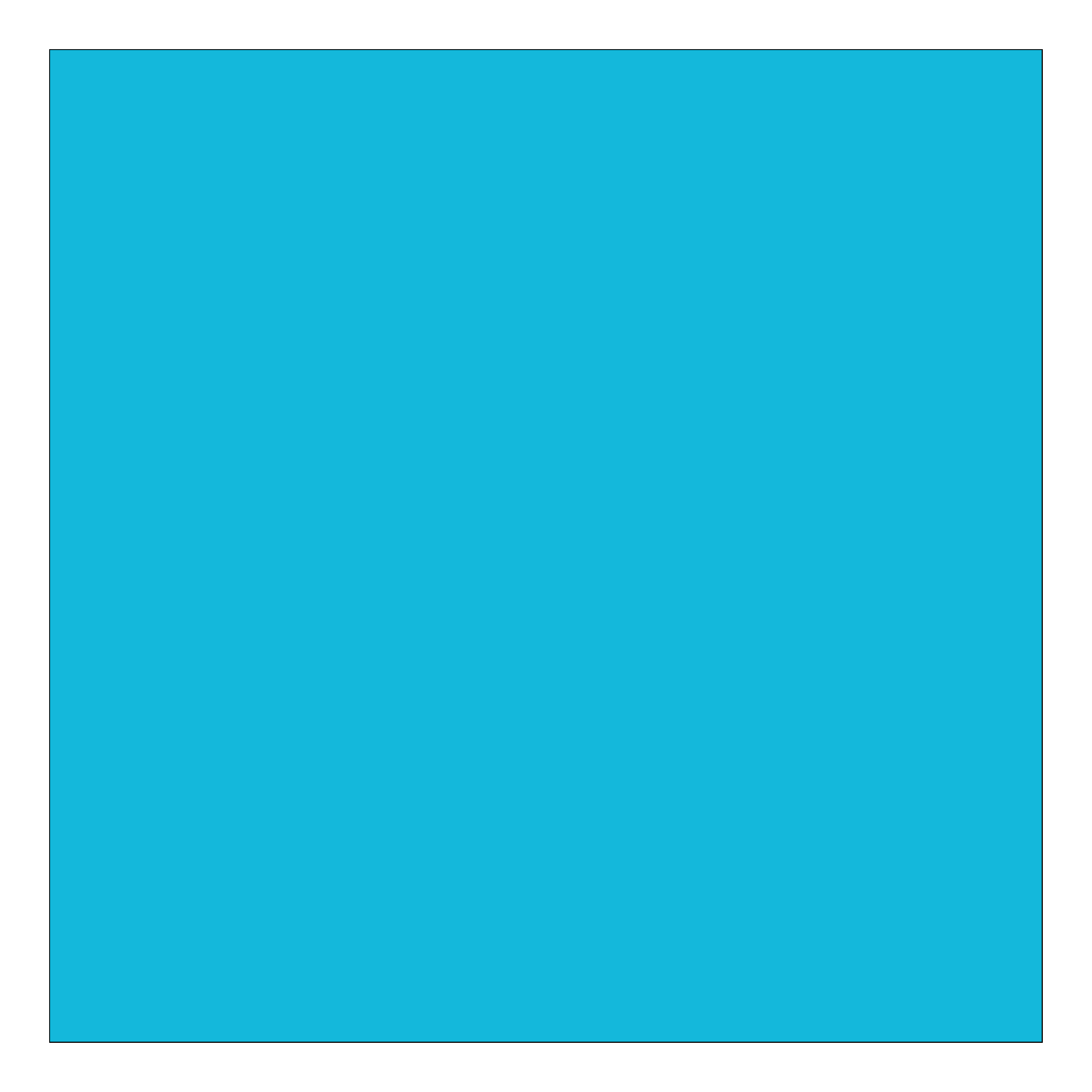}
  \includegraphics[height=2cm]{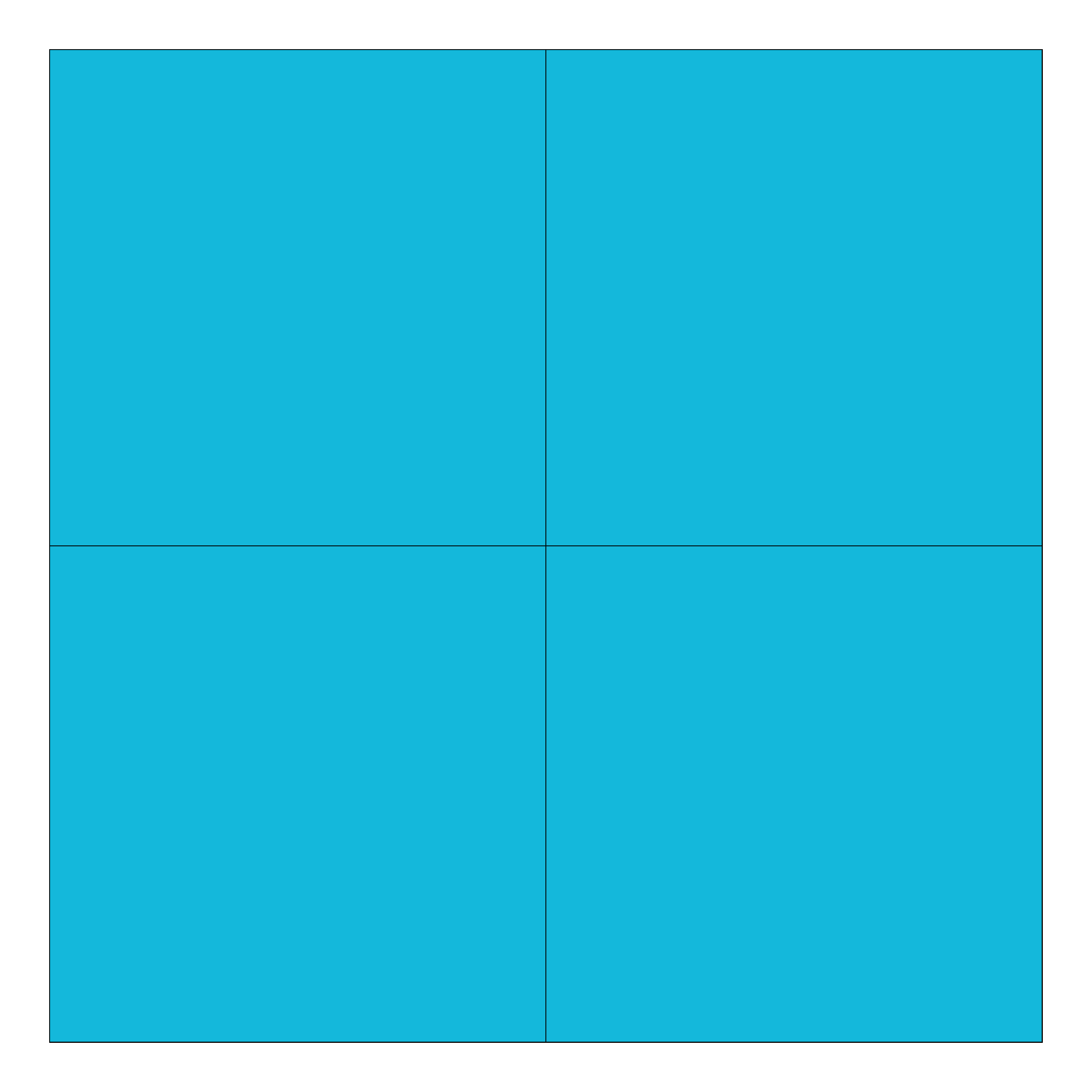}
  \includegraphics[height=2cm]{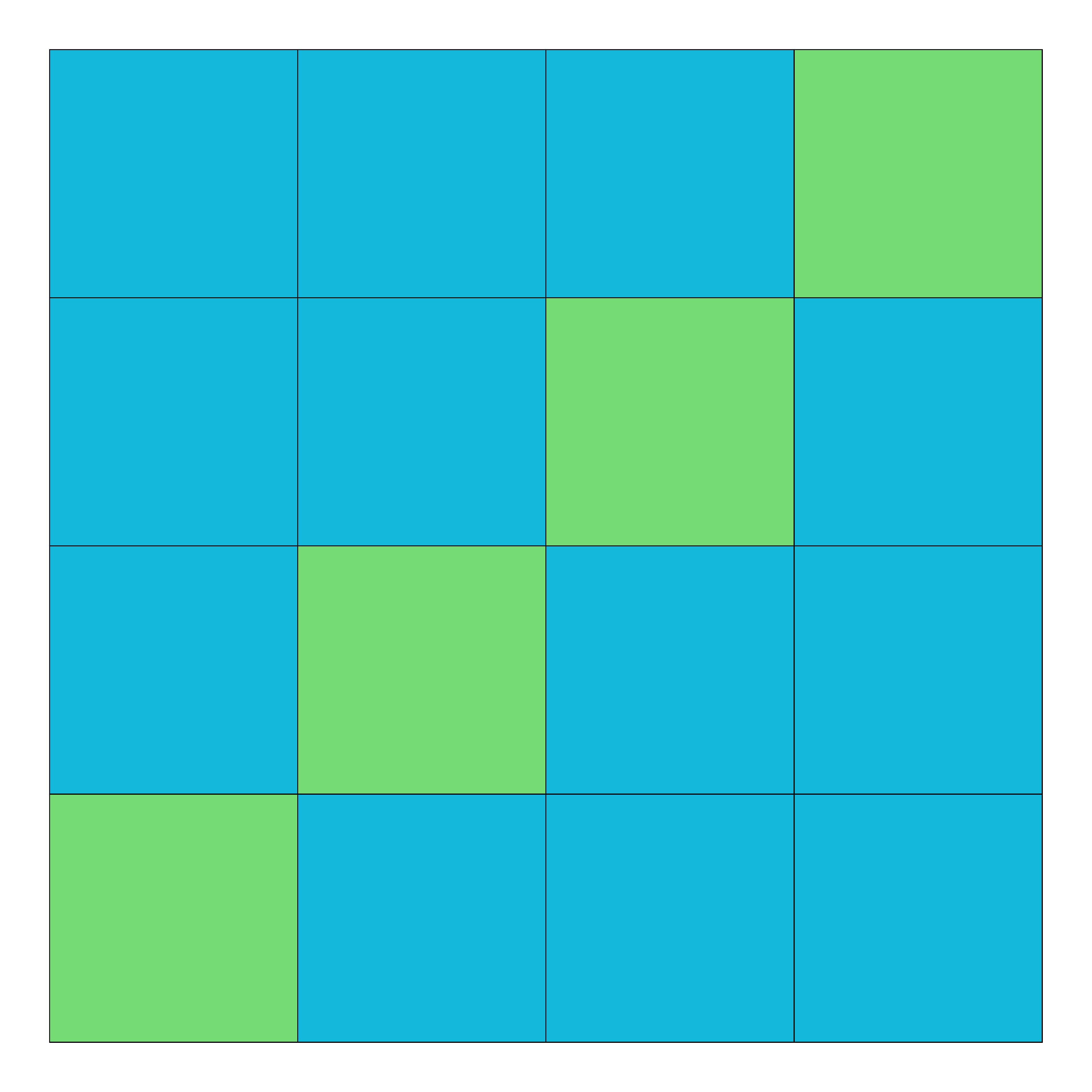}
  \includegraphics[height=2cm]{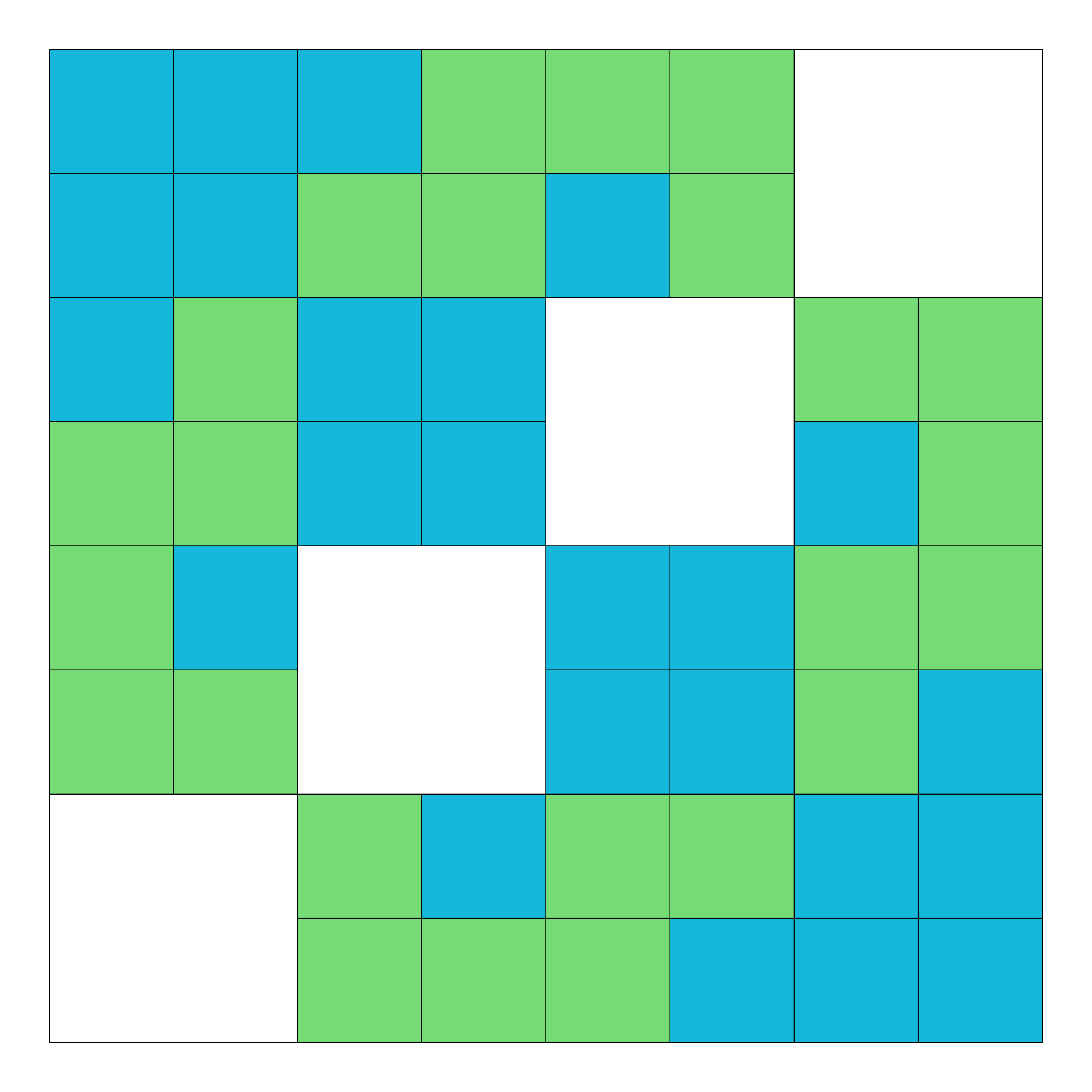}
  \includegraphics[height=2cm]{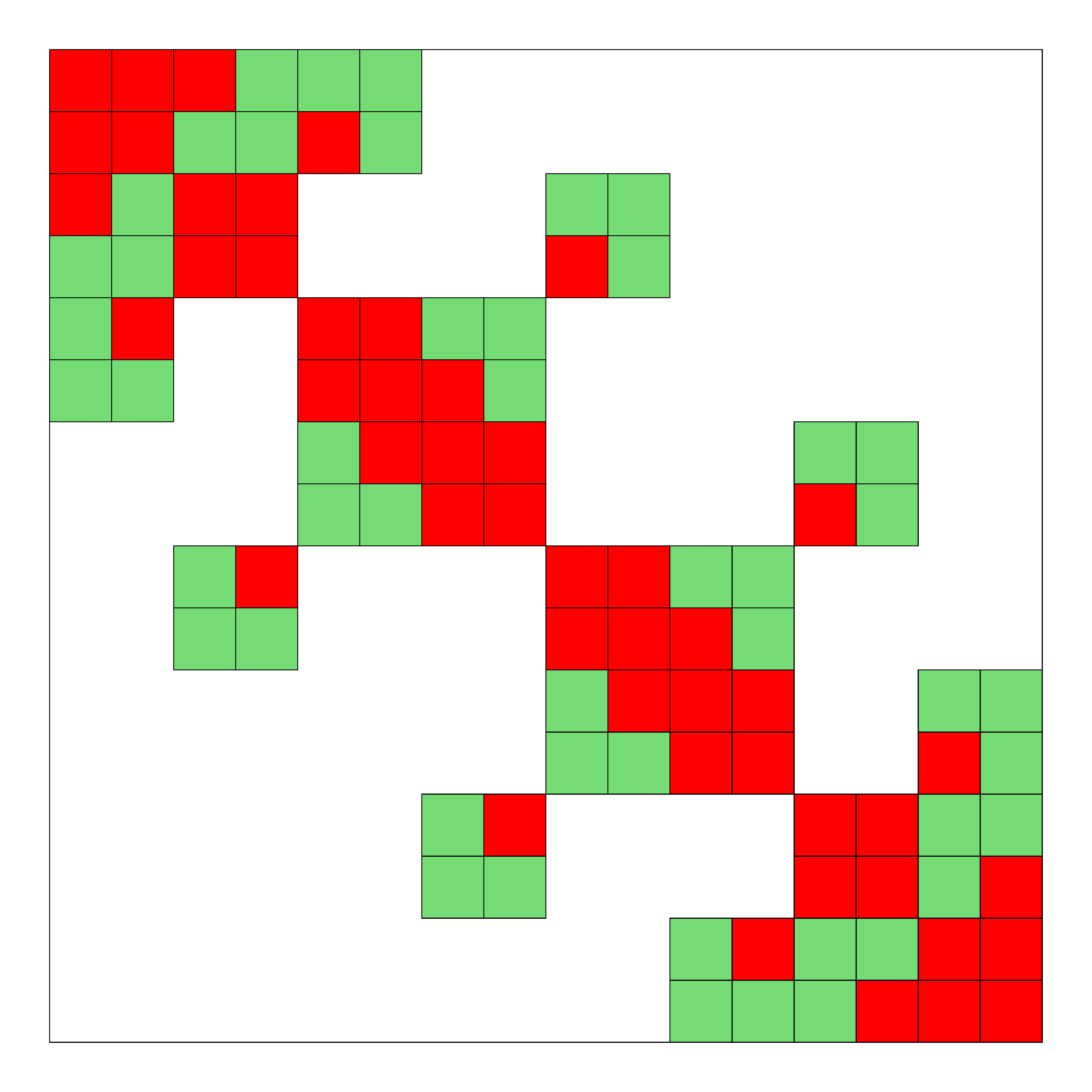}
  \vspace*{-5pt}
  \caption{Levels of the matrix tree. Inner nodes are in cyan, low rank leaves in green, and dense leaves in red.}
  \label{fig:hmatrix_levels}
\end{subfigure}
\caption{General partitioning of a hierarchical matrix.}
\label{fig:hmatrix_levels_and_leaves}
\end{figure}

\emph{Hierarchy of Blocks.}
Let $\mathcal{I}$ and $\mathcal{J}$ be the sets of indices of the rows and columns of a matrix, and $T_{\mathcal{I}}$ and $T_{\mathcal{J}}$ be hierarchical clusterings of these index sets, respectively. 
All blocks within the matrix can then be defined by cluster pairs $(t, s) \in T_{\mathcal{I} \times \mathcal{J}}$. Define an admissible block as a block that is either small enough to be stored in dense form of size $m \times m$, with $m$ denoting the so-called leaf size, or can be well approximated by a low rank matrix. A matrix $A$ in the $\mathcal{H}$-variant of hierarchical matrices can be partitioned into admissible blocks of various sizes that are organized hierarchically as the leaves of a \emph{matrix tree}. 
A low rank matrix $A^l_{ts}$ on level $l$ of the tree and defined by the cluster pair $(t, s)$ is represented as the outer product of two tall matrices $B^l_{ts}$ and $C^l_{ts}$ of rank $k^l$: $A^l_{ts} = B^l_{ts}{C^l_{ts}}^T$. 

The left panel of Figure \ref{fig:hmatrix_leaves} shows the leaves of this matrix tree which define a complete partitioning of the matrix into blocks, with dense leaves colored red and low rank leaves colored green. Figure \ref{fig:hmatrix_levels} shows the various levels of the tree, where inner nodes are in cyan. The structure of this matrix tree depends on the application. For example, in problems involving a spatial discretization, we can leverage data from the particles or nodes generated by the discretization to generate it with the aid of a geometric admissibility condition \cite{hackbusch15}. Other applications might rely on the available algebraic data or heuristics to determine which blocks of the matrix are admissible as low rank blocks. 


\emph{Hierarchy of Bases.} The \Htwo{} representation uses block low rank factorizations of the form $A^l_{ts} = U^l_t S^l_{ts} {V_s^{l}}^T$ for every $ts$ block at every level $l$, where $U^l_t$ and $V^l_s$ are bases for the block rows $t$ and block column $s$ of level $l$, respectively. $S^l_{ts}$ is a small $k_t^l \times k_s^l$ coupling matrix representing the interaction between the row and column cluster pairs. These coupling matrices are organized in the matrix tree $\cal{S}$ defined by $T_{\mathcal{I} \times \mathcal{J}}$. In our implementation, we currently use a fixed rank per level that we refer to as $k^l$, or even simply $k$ when the context is clear, which allows us to use higher-performing fixed-size batch linear algebra kernels when executing basis operations.


\begin{figure}
	\centering
  \vspace*{-5pt}
    \includegraphics[width=\linewidth]{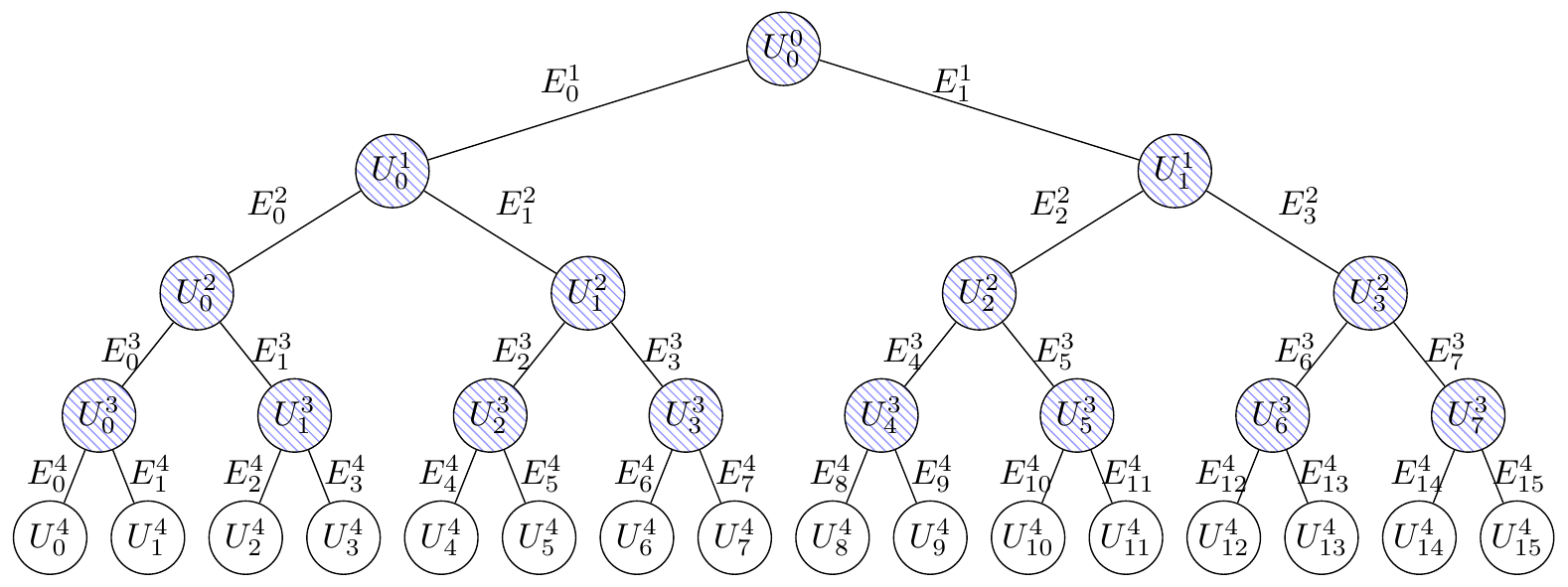}
    \caption{Basis tree $\cal{U}$ of an $\mathcal{H}^2$-matrix. Leaf nodes are stored explicitly, whereas inner nodes are represented implicitly using interlevel transfer matrices $E$. }
    \label{fig:hmatrix_basis}
\end{figure}

An additional hierarchy is introduced in the row and column basis trees, where a basis node is expressed in terms of the bases of its children. Basis nodes are only stored explicitly at the leaves of the tree, and inner nodes can be computed from their children using small $k^l \times k^{l-1}$ interlevel transfer matrices.
For example, one can compute an inner node $U^{l-1}_i$ from its two children, $U^l_{i_1}$ and $U^l_{i_2}$, and their corresponding transfer matrices $E^l_{i_1}$ and $E^l_{i_2}$ as
\begin{equation*}
\label{eq:transfer}
	U^{l-1}_i = 
	\begin{bmatrix} U^l_{i_1} & \\ & U^l_{i_2} \end{bmatrix}
	\begin{bmatrix} E^l_{i_1} \\ E^l_{i_2} \end{bmatrix}. 
\end{equation*}
The explicit bases at the leaves and the interlevel transfer matrices are organized in a \emph{basis tree}. Figure \ref{fig:hmatrix_basis} shows an example of the nested basis $\cal{U}$, with the implicitly represented inner nodes shaded.

Putting it all together, the $\mathcal{H}^2$-matrix representation may be succinctly written as 
\[ A = A_{de} + A_{lr} = A_{de} + <{\cal U}, {\cal S}, {\cal V}^T \!> \] 
where $A_{de}$ is a block sparse matrix with dense blocks of size $m \times m$, shown as the red leaves at the finest level of the matrix tree in Figure \ref{fig:hmatrix_leaves}.
$\cal{S}$ is a matrix tree of $k^l \times k^l$ coupling matrices, whose leaves, which appear at different levels, are shown as the cyan blocks in the right panel of Figure \ref{fig:hmatrix_leaves}, and $\cal{U}$ and $\cal{V}$ are the block row and block column basis trees, each 
consisting of $m \times k$ explicitly stored bases $U, V$ at the leaf level, and $k^l \times k^{l-1}$ interlevel transfer matrices $E$ and $F$, as shown in Figure \ref{fig:hmatrix_basis} for basis tree $\cal{U}$.
The triple-tree product notation $<{\cal U}, {\cal S}, {\cal V}^T\!>$ is defined as the assembly of all blocks $ts$ of all levels $l$ where every block may be computed as $U^l_t S^l_{ts} {V_s^{lT}}$.


\subsection{Distributed Representation and Construction}
\label{sec:dist_construction}

\begin{table} 
\begin{tabular}{l l} 
	\hline 
	Symbol & Description \\  \cline{1-1} \cline{2-2}
	%
	$P$, $p$ & Total number of GPUs, and index of the local GPU \\ 
	$\mathcal{U},\mathcal{V}$ & Complete basis trees \\ 	
	$\mathcal{S}$ & Complete matrix tree \\
	$\pp{\mathcal{U}}{p}$ &  Local branch of the basis tree on GPU $p$ \\ 
	$\pp{\mathcal{U}}{r}$ &  Root branch of the basis tree on the master process \\ 
	$\pp{\mathcal{S}}{p}$ &  Local branch of the matrix tree on GPU $p$ \\
	$\pp{\mathcal{S}}{r}$ &  Root branch of the matrix tree on the master process \\
	$\pp{\mathcal{S}}{p}_q$ &  Matrix tree branch, from dual tree traversal of $\pp{\mathcal{U}}{p}$ and $\pp{\mathcal{U}}{q}$\\
  $\batch{E}$, $\batch{x}$, etc. &  Marshaled $E$, $x$, etc.~arrays for batched execution \\
	$x, y$ &  Input and output vectors \\   
	$\pp{\widehat{x}}{p}, \pp{\widehat{y}}{p}$ &  Local branches of vector trees $\widehat{x}$ and $\widehat{y}$ on GPU $p$ \\
	$\pp{x}{p}, \pp{y}{p}$ &  Local sub-vectors of the input $x$ and output $y$ on GPU $p$ \\
	\hline 
\end{tabular}
\caption{Notation used. \vspace*{-6pt}}
\label{table:dist_notation} 
\end{table}

\begin{figure}[b]
	\centering
  \vspace*{-6pt}
	\includegraphics[width=0.9\columnwidth]{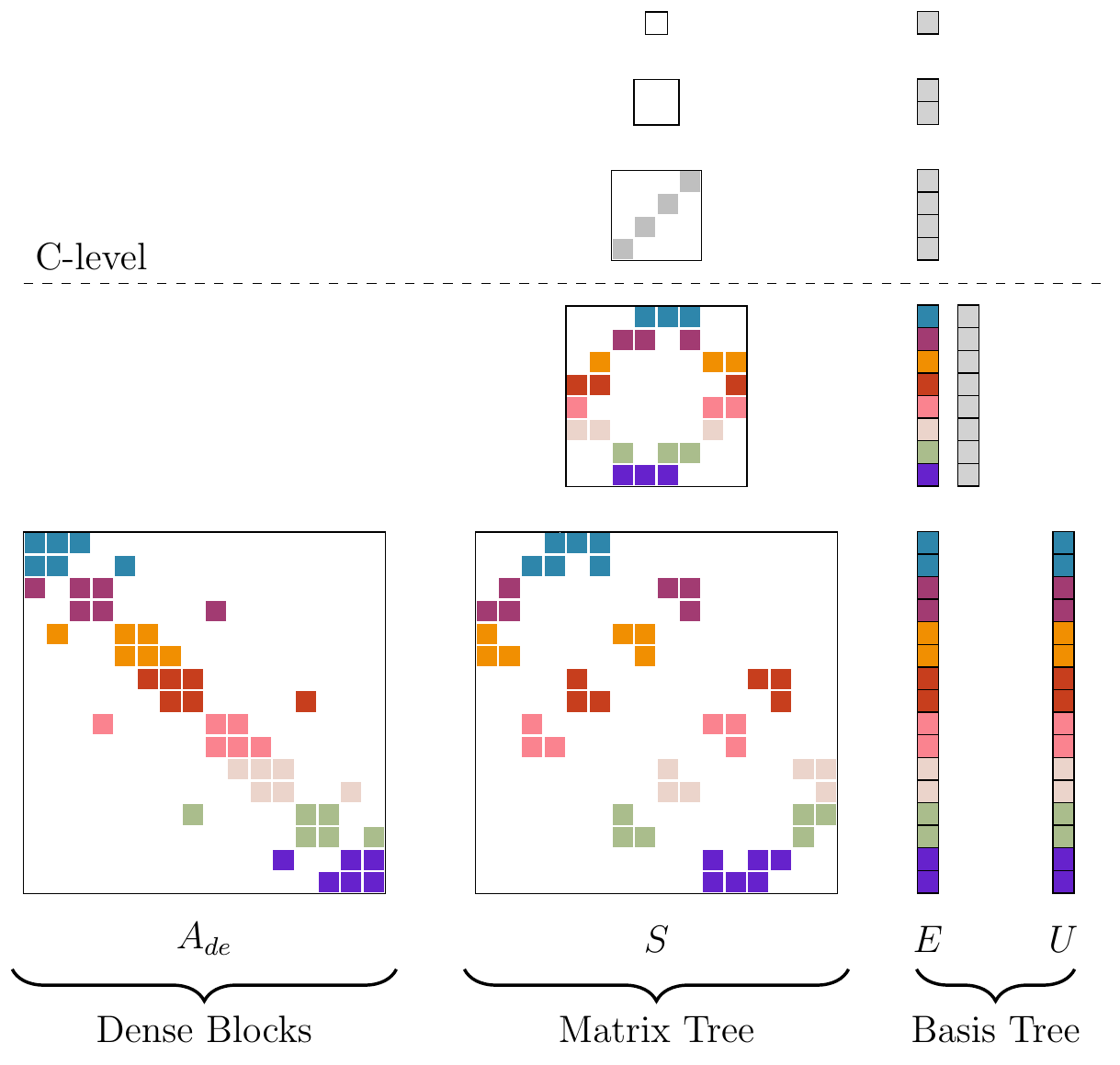}
	\caption{A hierarchical matrix distributed on 8 GPUs with the block rows of the dense and coupling leaves having the same color residing on the same GPU. Root branch is in grey.   \vspace*{-6pt} }
	\label{fig:distrbuted_hmatrix}
\end{figure}

The description of the parallel distribution of the matrix and the algorithms operating on it uses the notation in Table \ref{table:dist_notation}. We use a left subscript to refer to local data owned by a GPU, and a triple bar subscript to refer to data that has been marshaled to allow batched GPU kernels to be invoked on it. 

Treating each level of the matrix tree of the hierarchical matrix as a block sparse matrix, we decompose the levels into block rows, with each block row stored on a single GPU, as illustrated in Figure \ref{fig:distrbuted_hmatrix}.  The basis trees are similarly decomposed by level, assigning the nodes corresponding to the stored block rows/columns to the same GPU. This decomposition allows much of the upsweep and downsweep to be carried out independently as described in Section \ref{sec:dist_hgemv} for matrix-vector multiplication and in Section \ref{sec:compression} for matrix compression. Above a critical C-level, the decomposition stops and a single root GPU owns all top levels. 
The interlevel transfer operators at the roots of each of the local branch basis trees $\pp{\mathcal{U}}{p}$ are duplicated at the leaf level of $\pp{\mathcal{U}}{r}$ (see Figure \ref{fig:distrbuted_hmatrix}) to allow upsweep and downsweep operations to begin and end at the C-level. 

We construct the distributed matrix by first splitting the row cluster tree $T_\mathcal{I}$ into $P$ independent branches at level $l = \log P$ and a local branch $\pp{\mathcal{U}}{p}$ of the basis tree on GPU $p$ is generated for each $p \in \left\lbrace 0 \ldots P-1 \right\rbrace$. The top $l$ levels of the basis tree are kept on a master process and could potentially be split further if $P$ becomes large enough. 
The root branch of the matrix tree $\pp{\mathcal{S}}{r}$ can be generated on the master process in any number of ways, such as general admissibility dual tree traversal \cite{hackbusch15}.  
Once this matrix tree is constructed, a list of the basis nodes $L_p$ from $\pp{\mathcal{U}}{r}$ can be extracted for each GPU $p$ corresponding to the inadmissible nodes of the block row $p$ at level $l-1$ of the matrix tree. As an example, consider the distribution on $P=8$ GPUs of the fourth level of the matrix tree in Figure \ref{fig:hmatrix_levels}. The column indices corresponding to the cyan nodes are extracted for GPU $p=1$, setting $L_1=[0,1,4]$ from the second block row.
Once the list for each process has been compiled, they are scattered from the master process to all other processes.

Since the nodes $q$ in $L_p$ were generated from the inadmissible nodes of the matrix tree $S^{l-1}_{pq}$ that have to be subdivided further, the structure of the matrix tree $\pp{\mathcal{S}}{p}$ on each GPU can then be generated independently using multiple dual tree traversals of the root node of $\pp{\mathcal{U}}{p}$ with the nodes $q$ in $L_p$, where each traversal generates a local branch $\pp{\mathcal{S}}{p}_q$. 
Once the structure has been determined, the entries of the transfer and leaf matrices of the basis trees, together with the entries of the matrix tree can be populated independently on each GPU using established techniques \cite{borm10}.

\section{Distributed-memory Matrix-Vector Multiplication}
\label{sec:dist_hgemv}

\begin{figure}[t]
	\centering
	\includegraphics[width=\columnwidth]{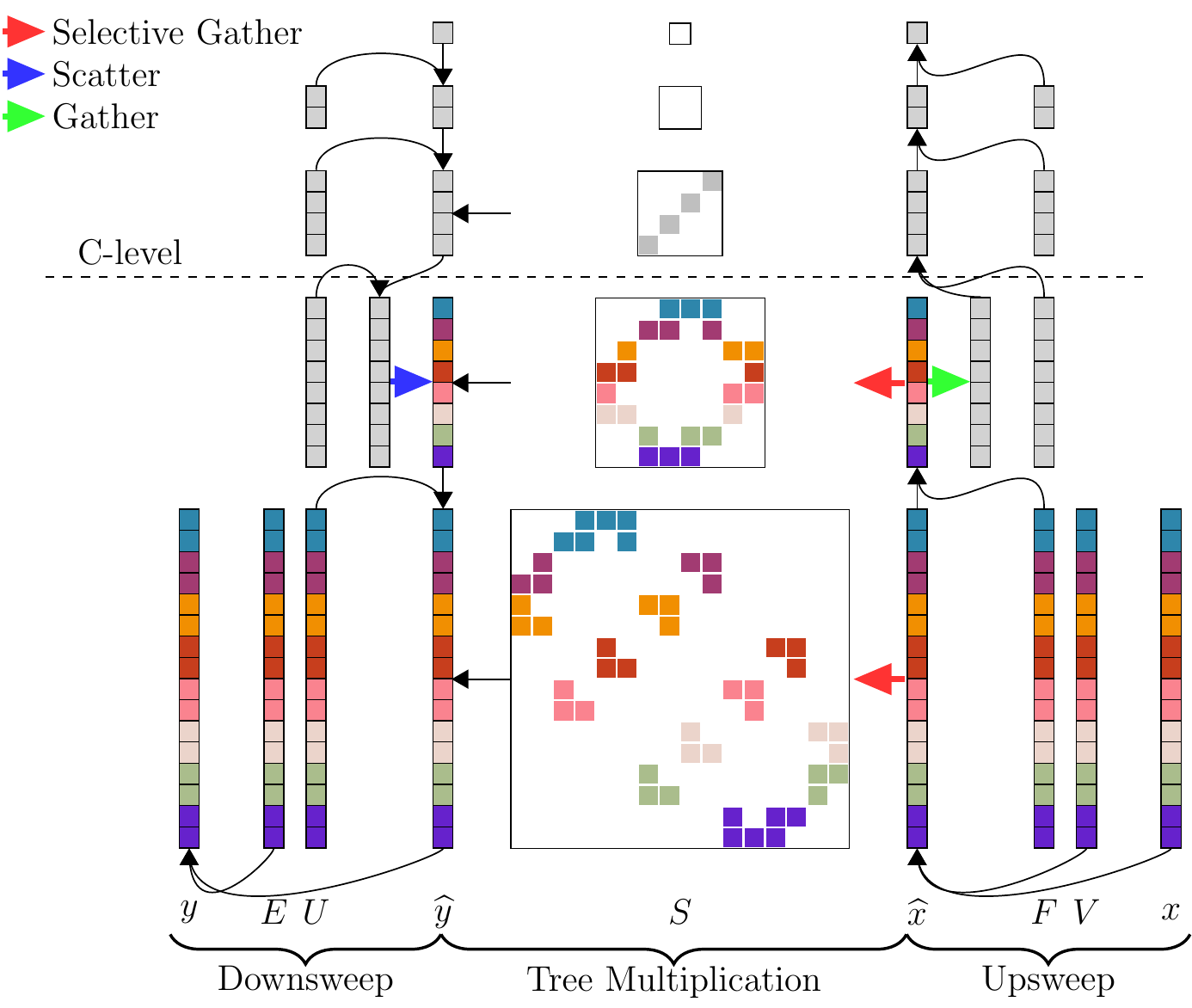}
	\caption{The three phases of the distributed vector product with the low rank part of the hierarchical matrix on 8 GPUs: upsweep, tree multiplication and downsweep. MPI communication is shown in the red, green, and blue arrows. Black arrows represent local operations. 
The gather of the root nodes of the branches of $\widehat{x}$ into the grey leaf nodes of $\pp{\widehat{x}}{r}$ is shown  as the green arrow. Similarly, the blue arrow shows the scatter of the leaves $\pp{\widehat{y}}{r}$. The red arrows indicate the selective gathers of the off-diagonal data from $\widehat{x}$ into each GPU.
  }
	\label{fig:distrbuted_hgemv}
\end{figure}

This section describes the overall structure of the distributed multiplication operation with $nv$ vectors concurrently. We use this structure to highlight the interprocess communication bottlenecks, that are then optimized in Section \ref{sec:dist_hgemv_opt}. 

The input multi-vector $x$ of size $N\times nv$ is distributed among $P$ GPUs, where each sub-vector $\pp{x}{p}$ is extracted from the index set defined by the root node of $\pp{\mathcal{U}}{p}$. The hierarchical matrix-vector product operation requires a standard block sparse multiplication $A_{de} x$ for the dense blocks of the matrix, which can be overlapped with the low rank multiplication $A_{lr} x = <\!{\cal U}, {\cal S}, {\cal V}^T\!\!> x$. The low rank portion of the product is a generalization to the hierarchical tree setting of a regular dense low rank matrix $USV^T$ by a vector, and is illustrated in Figure \ref{fig:distrbuted_hgemv}. In a first phase, we perform an upsweep of the basis tree ${\cal V}$ and compute $\widehat{x} = {\cal V}^T x$, which contains the products of the matrices $V_s^{lT}$ of all block columns $s$ at all levels $l$ with the corresponding pieces $x_s$ of $x$. In a second phase, a tree $\widehat{y} = {\cal S} \widehat{x}$ is computed. $\widehat{y}$ also consists of multilevel data corresponding to the products of $\widehat{y}^l = S^l \widehat{x}^l$ for every level $l$. In the third phase, a downsweep phase through the basis tree ${\cal U}$ accumulates the multilevel $\widehat{y}$ tree data into the output vector $y$, an operation we may symbolically write as $y = {\cal U} \, \widehat{y}$.


%

\subsection{Distributed Upsweep}
\begin{figure}
  \includegraphics[width=\columnwidth]{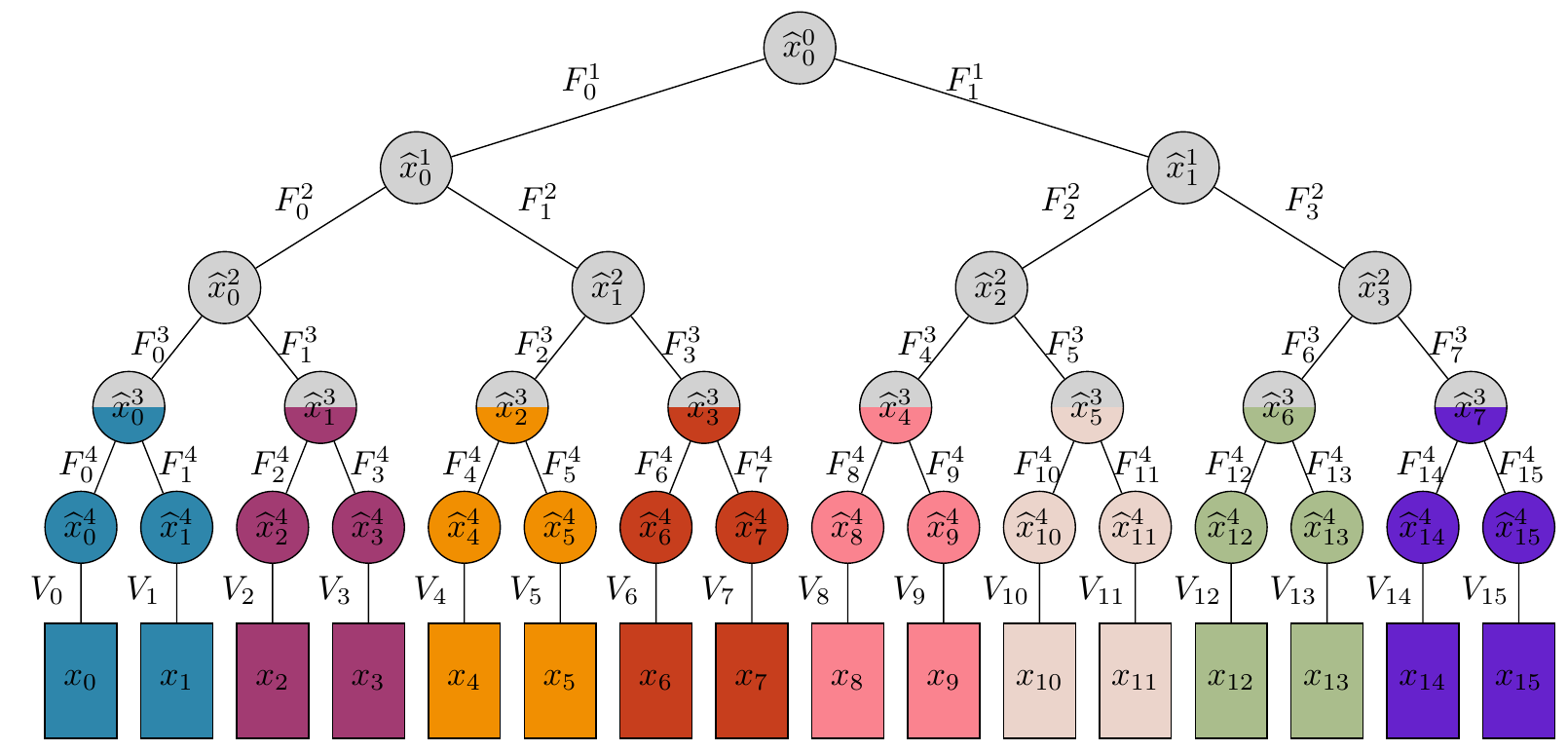}
  \caption{Distributed Upsweep \vspace*{-10pt}}
  \label{fig:dist_upsweep}
\end{figure}

The upsweep phase is illustrated in Figure \ref{fig:dist_upsweep} 
and summarized in Algorithm \ref{alg:naive_dist_upsweep}. It proceeds from the leaf level where the explicitly stored
block row bases, each of size $m \times k$, are simply multiplied by $x$. At every higher level, $\widehat{x}^{l-1}$ can be computed from the children at level $l$ using the transfer operators \cite{borm10}. For a parent node $s$ with children $s_1$ and $s_2$ we have
\[
\widehat{x}^{l-1}_{s} = 
\begin{bmatrix}
F_{s_1}^{l\,T} & {F_{s_2}^{l\,T}} 
\end{bmatrix}
\begin{bmatrix}
{V_{s_1}^{l\,T}} &     \\
& {V_{s_2}^{l\,T}} \\
\end{bmatrix}
\begin{bmatrix}
x_{s_1}\\
x_{s_2}
\end{bmatrix} = 
{F_{s_1}^{lT}} \widehat{x}^{l}_{s_1} + {F_{s_2}^{lT}} \widehat{x}^{l}_{s_2}.
\]

\begin{algorithm}
\caption{Single GPU Upsweep for forming local $\pp{\widehat{x}}{p}$}
\label{alg:upsweep2}
\begin{algorithmic}[1]
\Procedure{upsweep}{$V$, $F$, $x$, $\widehat{x}$}
	\State $q$ = depth$( \widehat{x} )$     \Comment{{\small \emph{leaf level, log$(\pp{N}{p}/m)$}}}
	\State $\widehat{x}^{q} = $ gemvBatched $(\pp{N}{p}/m, V^T, x)$ \Comment{{\small \emph{$\widehat{x}^{q}=V^T x$}}}
	\For{$l$ = $q \rightarrow {}1$} 	\Comment{{\small \emph{up the $\cal{V}$ tree}}}
		\State $nb$ = $\pp{N}{p} / m / 2^{q-l+1}$
		\State $[\batch{F^l}, \batch{\widehat{x}^l}, \batch{\widehat{x}^{l-1}}]$ = marshalUpsweep$(F^{l}, \widehat{x}^{l}, \widehat{x}^{l-1})$
		\vspace{0pt}
		\For{$j$ = $1 \rightarrow 2$} 		\Comment{{\small \emph{binary tree}}}
			\State $\batch{\widehat{x}^{l-1}}(j)\mathrel{+}= $ gemvBatched $(nb, \batch{F^l}(j)^T, \batch{\widehat{x}^l}(j))$
		\EndFor
	\EndFor
\EndProcedure
\end{algorithmic}
\end{algorithm}

\begin{algorithm}
\caption{Distributed Upsweep}
\label{alg:naive_dist_upsweep}
\begin{algorithmic}[1]
\Procedure{DistUpsweep}{$P$, $p$, $\pp{x}{p}$, $\pp{\mathcal{V}}{r}$, $\pp{\mathcal{V}}{p}$, $\pp{\widehat{x}}{r}$, $\pp{\widehat{x}}{p}$}
  	\State upsweep$(\pp{\mathcal{V}}{p}.V, \pp{\mathcal{V}}{p}.F, \pp{x}{p}, \pp{\widehat{x}}{p} )$
 	\If $p = 0$
    \vspace*{-1pt}
		\State $l = $ depth$( \pp{\mathcal{V}}{r} )$ \Comment{{\small \emph{C-level}}}
    \LineComment{{\small \emph{Gather the roots of $\pp{\widehat{x}}{p}$ into the leaf level of  $\pp{\widehat{x}}{r}$}}}
    \vspace*{-0.5pt}
		\State $\pp{\widehat{x}^l}{r} = $ gather$( \pp{\widehat{x}^0}{p} )$ 
    \LineComment{{\small \emph{Ignore the leaves by passing \texttt{null}}}}
    \vspace*{-0.5pt}
		\State upsweep$\left( \text{null}, \pp{\mathcal{V}}{r}.F, \text{null}, \pp{\widehat{x}}{r} \right)$ 
	\EndIf
\EndProcedure
\end{algorithmic}
\end{algorithm}

In the distributed setting, the upsweep on each GPU $p$ can proceed in parallel on each of the local branches $\pp{\mathcal{V}}{p}$ independently, using the kernel shown in Algorithm \ref{alg:upsweep2}. Once the upsweep reaches the roots of each branch, the data from all root nodes of the $\pp{\widehat{x}}{p}$ trees are gathered on the master process to populate the leaf level of the root branch $\pp{\mathcal{V}}{r}$, allowing the upsweep to complete. The regular upsweep Algorithm \ref{alg:upsweep2} can be used for the branch upsweep as well as the root upsweep by simply omitting the first batched operation on the leaves of the root branch, since that step is replaced by the gathering of the data.

From an efficiency perspective, the primary challenge in the upsweep comes from the fact that the individual operations in the basis tree nodes involve small matrix operators that do not possess a sufficient amount of data parallelism for effective GPU utilization. We overcome this problem by marshaling the appropriate level data to allow batched GPU kernels to be invoked, allowing $\widehat{x}$ to be computed with only $O(\log N)$ batched kernel executions. Except for the top few levels of the root process $r$, these executions happen concurrently on the $P$ GPUs. The marshaling GPU kernel is shown for a step of the upsweep operation in Algorithm \ref{alg:hgemv_marshalupsweep}. The marshaling kernel essentially plays the role of a fast static scheduler for all operations performed on a given level $l$ of the tree. It prepares for the execution of the operations by efficiently gathering data from the basis tree. Marshaling involves no data movement and therefore has very little overhead and in practice consumes only a tiny fraction of total execution time. 

\begin{algorithm}
\caption{GPU Upsweep Marshaling Kernel}
\label{alg:hgemv_marshalupsweep}
\begin{algorithmic}[1]
\Procedure{marshalUpsweep}{$F^{l}$, $\widehat{x}^{l}$, $\widehat{x}^{l-1}$}
  \vspace*{-2pt}  
  \LineComment{{\small \emph{uses the arrays of a flattened tree data structure }}}
	\State $n_p$ = levelptr$[l-1]$, $k_p$ = levelrank$[l-1]$
	\State $n_c$ = levelptr$[l]$, $k_c$ = levelrank$[l]$	
	\ForAllp{$p = n_p \rightarrow n_c$} 
    \vspace*{-1pt}
		\State $i = p - n_p$     \Comment{{\small \emph{Batch index}}}
		\State $c$ = head$[p]$, $c_i = 0$
		\While {$c \neq \text{empty}$}
      \vspace*{-1pt}  
      \LineComment{{\small \emph{Extract level pointer data}}}
			\State $\batch{F^l}(c_i)[i] = $ ptr$(F^{l}) + (c - n_c) \times k_c \times k_p$ 
			\State $\batch{\widehat{x}^l}(c_i)[i] = $ ptr$(\widehat{x}^{l}) + (c - n_c) \times k_c$
			\State $\batch{\widehat{x}^{l-1}}(c_i)[i] = $ ptr$(\widehat{x}^{l-1}) + i \times k_p$
			\State $c = $ next$[c]$, $c_i = c_i + 1$ 
		\EndWhile
	\EndForAllp
\EndProcedure
\end{algorithmic}
\end{algorithm}

\vspace*{-4pt} 
\subsection{Distributed Intermediate Multiplication}
The second phase of the operation builds a vector tree $\widehat{y}$, where each node $t$ in a level $l$ is the product of the block row $t$ of level $l$ of the coupling matrix tree with the corresponding nodes in $\widehat{x}$. This operation can be expressed as
\[ \vspace*{-3pt} \widehat{y}_t^{l} = \sum_{s \in \{b_t\}} S_{ts}^l \widehat{x}_s^{l} \] 
where $b_t$ is the set of column indices of the matrix blocks in the block row $t$. This is a block sparse matrix-vector multiplication at every level, and is illustrated in the middle portion of Figure \ref{fig:distrbuted_hgemv}, where the block-row, multi-GPU decomposition of each level is also highlighted.
The scalar version of this problem is a well-studied computational kernel that has many possible high quality solutions \cite{guo16,merrill16,bienz19,elafrou19} which may be adapted to the block-sparse version. While we could rely on vendor kernels, their relatively low performance and lack of support for non-uniform blocks necessitates a different approach. Our solution relies on a key result regarding hierarchical matrices which puts a bound on the sparsity constant $C_{sp}$, the maximum number of blocks in any block row $t$ at any level $l$ of the matrix tree. Such constant is bounded by a dimension-independent $O(1)$ value that only depends on the specific structure of the matrix and the admissibility criterion\cite{grasedyck03,borm10}.

During the construction of the matrix tree, we generate conflict-free batches of matrix products that can then be executed by a series of non-uniform batched matrix-vector and matrix-matrix kernels. A conflict-free ordering of the batch indices can be obtained by assigning a batch index based on the position within the block row or column that increases from left to right, allowing us to marshal all the batches in a single marshaling kernel. While there will potentially be some kernels that perform little work, the bounded sparsity constant guarantees that there will be few of them, and they will thus represent a small portion of the total runtime. This could be optimized further by running those small batches on a separate low priority stream and then combining the results with the main execution stream.

The boundedness of the sparsity constants $C_{sp}$ also guarantees that the block row local to a GPU will require input from $\widehat{x}$ nodes belonging to a limited number of remote GPUs: we therefore consider a standard approach in distributed memory sparse matrix vector products, and split the block row into diagonal and off-diagonal parts. The contributions of these two parts can be overlapped with the needed communication as described in Section \ref{sec:fastgather}. Once the needed parts of $\widehat{x}$ are assembled, the multiplication phase can proceed on each GPU independently on the coupling matrices of $\pp{\mathcal{S}}{p}$ using the \texttt{treeMultiply} routine of Algorithm \ref{alg:mult}. Processing the multiplication of the root branch $\pp{\mathcal{S}}{r}$ on the master process to produce the root branch $\pp{\widehat{y}}{r}$ finalizes the phase as shown in Algorithm \ref{alg:naive_dist_mult}. The \texttt{selectiveGather} routine communicates and assembles the needed portion of the $\widehat{x}$ tree from the remote branches to use in the local \texttt{treeMultiply}. This is described in Section \ref{sec:fastgather}.


\newlength{\textfloatsepsave} \setlength{\textfloatsepsave}{\textfloatsep} 
\setlength{\textfloatsep}{-20pt}

\begin{algorithm}
\caption{GPU Matrix Tree Multiplication for $\widehat{y}$}
\label{alg:mult}
\begin{algorithmic}[1]
\Procedure{treeMultiply}{$S$, $\widehat{x}$, $\widehat{y}$} 
	\State $q$ = depth$( \widehat{y} )$   
	\ForAllp{ $l$ = $1 \rightarrow q$}
		\State $\widehat{y}^{l}$ = blockSparseMV$( S^{l}, \widehat{x}^{l})$ \Comment{{\small \emph{conflict-free batches}}}
	\EndForAllp
\EndProcedure
\end{algorithmic}
\vspace*{-2pt}
\end{algorithm}

\begin{algorithm}
\caption{Distributed Multiplication}
\label{alg:naive_dist_mult}
\begin{algorithmic}[1]
\Procedure{DistTreeMult}{$P$,$p$, $\pp{\widehat{x}}{r}$, $\pp{\widehat{x}}{p}$, $\pp{\mathcal{S}}{r}$, $\pp{\mathcal{S}}{p}$, $\pp{\widehat{y}}{r}$, $\pp{\widehat{y}}{p}$}
  	\State $\widehat{x} = $ selectiveGather$( \pp{\widehat{x}}{p}, P )$ \Comment{{\small \emph{See section} \ref{sec:fastgather} }}
  	\State treeMultiply$\left( \pp{\mathcal{S}}{p}, \widehat{x}, \pp{\widehat{y}}{p} \right)$ 
  \vspace*{-1pt} 
 	\If $p = 0$
		\State treeMultiply$\left( \pp{\mathcal{S}}{r}, \pp{\widehat{x}}{r}, \pp{\widehat{y}}{r} \right)$ 
	\EndIf
\EndProcedure
\end{algorithmic}
\vspace*{-2pt}
\end{algorithm}

\setlength{\textfloatsep}{\textfloatsepsave}

\subsection{Distributed Downsweep}

The vector tree $\widehat{y}$ now contains, at every level $l$, a partial contribution to the output vector $y$ expressed in terms of the bases $U^l$, and it could be computed as $U_t^l \widehat{y}_t$ if the bases at level $l$ were explicitly available. However, since we only store interlevel transfer operators, we express $\widehat{y}_t^l$ in terms of bases of the children of node $t$ and accumulate them in a downsweep phase through the ${\cal U}$ tree from root to leaves. This partial accumulation takes the form
\[
U^{l-1}_{t} \widehat{y}^{l-1}_{t} + \begin{bmatrix}
U^l_{t_1} \widehat{y}^l_{t_1} \\
U^l_{t_2} \widehat{y}^l_{t_2}
\end{bmatrix} = 
\begin{bmatrix}
U^l_{t_1}  & 		\\
& U^l_{t_2} 
\end{bmatrix} \begin{bmatrix}
E^l_{t_1} \hat{y}^{l-1}_{t} + \widehat{y}^l_{t_1}\\
E^l_{t_2} \hat{y}^{l-1}_{t} + \widehat{y}^l_{t_2}
\end{bmatrix}
\]
between a parent node at level $l-1$ and children $t_1$ and $t_2$ at the finer level $l$. 

The downsweep procedure, pictured in the left part of Figure \ref{fig:distrbuted_hgemv}, is summarized in Algorithm \ref{alg:naive_dist_downsweep}.
A downsweep through the root subtree $\pp{\widehat{y}}{r}$ is first performed. Once the leaf level of $\pp{\widehat{y}}{r}$ has been updated, it can be scattered from the master process to all other GPUs and added to the roots of local trees $\pp{\widehat{y}}{p}$ containing data from the previous phase of the multiplication.
When the roots of the local branches $\pp{\widehat{y}}{p}$ on the $P$ GPUs have their proper accumulation, obtained from the leaves of the tree $\pp{\widehat{y}}{r}$, we can sweep down $\pp{\widehat{y}}{p}$ independently on all GPUs
setting each node to 
$\pp{\widehat{y}}{p}_{t_i}^{l} = \pp{\widehat{y}}{p}_{t_i}^{l} + \pp{E_{i}^{l}}{p} \ \pp{\widehat{y}}{p}_{t}^{l-1} $; the level $l$ at each step now also contains the partial sum of $\pp{y}{p}$ for all levels above $l$ expressed in the basis of $U^l$. The leaf level will then contain the complete sum which is finally expanded into $\pp{y}{p}$ through a multiplication by the explicitly stored leaf level bases. We follow the same approach used for the upsweep, where each level is also processed in parallel by first marshaling the operations and then executing using a batch matrix vector product. This leads us to Algorithm \ref{alg:downsweep} for computing $\pp{y}{p}$, i.e., the distributed vector $y$. The downsweep marshaling algorithm is structurally similar to the one described in Algorithm \ref{alg:hgemv_marshalupsweep} and is omitted for brevity.

\begin{algorithm}
\caption{Single-GPU downsweep for forming local $\pp{y}{p}$}
\label{alg:downsweep}
\begin{algorithmic}[1]
\Procedure{downsweep}{$U$, $E$, $\widehat{y}$, $y$}
	\State $q$ = depth$( \widehat{y} )$     \Comment{{\small \emph{leaf level, log($\pp{N}{p}/m)$}}}

	\For{$l$ = $1 \rightarrow q$} 	\Comment{{\small \emph{down the $\cal{U}$ tree}}}
		\State $nb$ = $\pp{N}{p} / m / 2^{q-l}$
		\State $[\batch{E^l}, \batch{\widehat{y}^{l-1}}, \batch{\widehat{y}^l}]$ = marshalDownsweep$(E^{l}, \widehat{y}^{l-1}, \widehat{y}^{l})$
		\State $\batch{\widehat{y}^{l}} \mathrel{+}= $ gemvBatched$(nb, \batch{E^l}, \batch{\widehat{y}^{l-1}})$
	\EndFor
	
	\State $y = $ gemvBatched $\left( \pp{N}{p}/m, U, \, \widehat{y}^{q} \right)$
\EndProcedure
\end{algorithmic}
\end{algorithm}


\begin{algorithm}
\caption{Distributed Downsweep}
\label{alg:naive_dist_downsweep}
\begin{algorithmic}[1]
\Procedure{DistDownsweep}{$P$, $p$, $\pp{y}{p}$, $\pp{\mathcal{U}}{r}$, $\pp{\mathcal{U}}{p}$, $\pp{\widehat{y}}{r}$, $\pp{\widehat{y}}{p}$}
 	\If $p = 0$ \Comment{{\small \emph{Ignore the leaves by passing \texttt{null}}}}
 		\State downsweep$( \text{null}, \pp{\mathcal{U}}{r}.E, \pp{\widehat{y}}{r}, \text{null})$ 
	\EndIf
	\State $l = $ depth$( \pp{\mathcal{U}}{r} )$ \Comment{{\small \emph{C-level}}}
  \LineComment{{\small Scatter the leaf level of $\pp{\widehat{y}}{r}$ into the roots of $\pp{\widehat{y}}{p}$}}
	\State $\pp{\widehat{y}^0}{p} = \pp{\widehat{y}^0}{p} + \text{scatter}( \pp{\widehat{y}^l}{r} )$ 
	\State downsweep$\left(\pp{\mathcal{U}}{p}.U, \pp{\mathcal{U}}{p}.E, \pp{y}{p}, \pp{\widehat{y}}{p} \right)$
\EndProcedure
\end{algorithmic}
\end{algorithm}

\section{Optimizing Communication}
\label{sec:dist_hgemv_opt}

\subsection{Optimizing Communication Volume}
\label{sec:fastgather}

We first discuss a communication-optimized parallel block sparse matrix-vector multiplication for a given level matrix $S^l$ of $A_{lr}$, the low rank portion of the hierarchical matrix; the same logic applies to $A_{de}$ and it is omitted for brevity. 

The matrix tree is first split into two distinct trees: a diagonal matrix tree $\pp{S^l}{p}_p$ and an off-diagonal matrix tree $\pp{S^l}{p}_{\comp{p}}$. While the off-diagonal portion could simply be represented as a set of trees, with one tree for every interaction $\pp{S^l}{p}_{q}$ with a GPU $q$, it is far more efficient to have them all in a single flattened one to allow marshaling operations on the off-diagonal blocks to be completed with a single kernel call.

\begin{figure}
  \vspace*{-12pt}
	\centering
	\includegraphics[width=0.8\columnwidth]{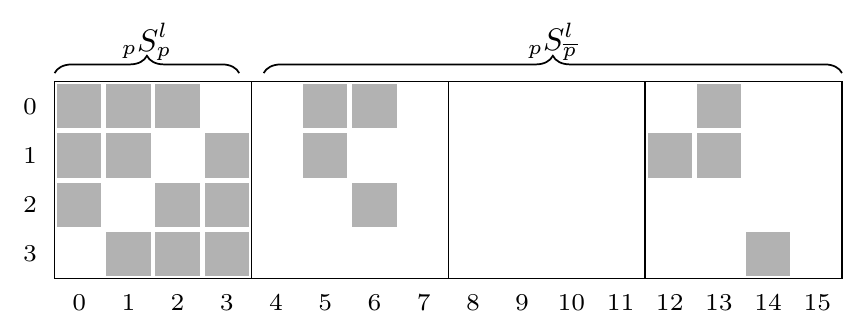}
	\def\arraystretch{1.0}
	\newcolumntype{C}{>{\centering\arraybackslash}p{2.5ex}}
	\begin{tabular}{|c|C|C|C|C|C|C|C|C|C|C|C|C|C|C|C|}
		\hline
		\texttt{pid}        & 1 & 3 & & & \\ \hline
		\texttt{nodes\_ptr} & 0 & 2 & 5 & & \\ \hline
		\texttt{nodes}      & 5 & 6 & 12 & 13 & 14 \\ \hline
	\end{tabular}
	\caption{The compressed node data for $\pp{S^l}{0}_{\comp{p}}$ of a hierarchical matrix distributed to 4 GPUs. Process 2 has no corresponding basis tree data and it is not listed in \texttt{pid}. \vspace*{-10pt}}
	\label{fig:compressed_vnodes}
\end{figure}

Once the trees are split, the list of basis nodes that interact with $\pp{S^l}{p}_{\comp{p}}$ can be generated by iterating through its $(t, s)$ pairs 
and determining all unique $s$ values. 
Given the boundedness of the sparsity constant $C_{sp}$,
on a given level $l$, a GPU $p$ will receive data only from a few other GPUs; we thus determine the list of GPUs that need to send data to $p$, as well as the list of nodes that should be received, and store such information in a compressed-storage format as representatively shown in Figure \ref{fig:compressed_vnodes}. The data needed from process \texttt{pid}$[i]$, corresponding to global indices listed in \texttt{nodes} from \texttt{nodes\_ptr}$[i]$ to \texttt{nodes\_ptr}$[i+1]$, is communicated among GPUs during the setup phase. A list of unique entries of \texttt{nodes} represents the compressed storage for the offiagonal block.

The diagonal multiplication phase can proceed without any communication while the off-diagonal phase needs input from other GPUs. The nodes at a level $l$ of the local $\pp{\widehat{x}}{p}$ branch which are needed by other GPUs are packed into a single buffer using a marshaling kernel and a single batched copy kernel populates a send buffer $B_s^l$, which is then used to issue non-blocking sends to the neighboring GPUs. 
Non-blocking receives are issued to populate a receive buffer $B_r^l$, which can then be directly used for the off-diagonal level of the tree, since the column basis indices have already been updated to use the compressed node format.

Algorithm \ref{alg:optimized_dist_mult} summarizes the overall communication-optimized multiplication phase. The \texttt{exchangeData} routine executes the non-blocking MPI sends and receives using the compressed off-diagonal data that is marshaled using the \texttt{marshalOffdiag} routine and copied using a batch kernel.


\begin{algorithm}
\caption{Optimized Multiplication Phase}
\label{alg:optimized_dist_mult}
\begin{algorithmic}[1]
\Procedure{optTreeMult}{$\pp{\mathcal{S}}{p}, \pp{\widehat{x}}{p}, \pp{\widehat{y}}{p}$, $\text{recv}$, $\text{send}$}
	\State $nv = $ vectors($\pp{\widehat{x}}{p}$)
	\State $d = $ depth($\pp{\widehat{x}}{p}$)
	\For $l = 0 \rightarrow d$
    	\State $[\batch{B}, \batch{x}]$ = \text{marshalOffdiag}$\left( B_s^l, \widehat{x}^l, \text{send}, nv \right)$
    	\State batchCopy $(\batch{B}, \batch{x})$
    	\LineComment{{\small \emph{non-blocking Isends and Irecvs}}}
		\State \text{exchangeData}($B_s^l$,$B_r^l$, $\text{recv}[l]$,$\text{send}[l]$,$nv$) 
	\EndFor
	\State treeMultiply$(\pp{\mathcal{S}}{p}_p, \pp{\widehat{x}}{p}, \pp{\widehat{y}}{p})$ \Comment{{\small \emph{Diagonal mult}}}
	\State waitAll(\,) \Comment{{\small \emph{Wait for all transfers to complete}}}  
  \label{alg:optimized_dist_mult:wait_all}
	\State treeMultiply$(\pp{\mathcal{S}}{p}_{\comp{p}}, B_r, \pp{\widehat{y}}{p})$ \Comment{{\small \emph{Off-diagonal part}}}
\EndProcedure
\end{algorithmic}
\end{algorithm}

\subsection{Achieving Communication and Computation Overlap}

While the communication volume has been reduced and a portion of the communication can be hidden by overlapping it with the diagonal multiplication, there are still a few challenges left that hold back performance. Assuming the lack of hardware support for more advanced memory transfer features such as GPUDirect RDMA, transferring the data from the GPU to the host adds some overhead to the execution, as well as synchronization points that impact GPU usage. Moreover, not all MPI distributions guarantee that communication can progress in the background without explicit intervention from the process.

The effectiveness of overlapping the diagonal multiplication phase with the needed communication depends on the structure of the tree, the capabilities of the communication network, and the compute capabilities of the GPU. For a tailored, fine-grained control of the communications involved in the algorithm at hand, we explicitly create communication threads that queue up asynchronous copies on separate streams to overlap the transfers with the processing of each level of the local branch upsweep. The transfers can then be carried out on the same thread without interrupting or forcing synchronizations with the main execution stream, allowing communication and computation to overlap. When the diagonal block kernels have been queued up on the stream, the communication thread can then join the main thread. The multiplication of the root branch on the master process can also be hidden, to some extent, by overlapping it with the main stream of work, and scheduling it on a low priority stream. This will allow the work to be completed during phases of low GPU utilization, such as the smaller top levels of the basis and matrix tree. Finally, by adding the dense block multiplication phase to a low priority stream, GPU utilization on all GPUs can be increased as well.

\begin{figure*}[ht]
	\setlength{\fboxsep}{0pt}
	\raggedright \fbox{\includegraphics[height=.72cm]{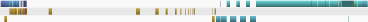}}
	\\
	\vspace*{2pt}
	\raggedright \fbox{\includegraphics[height=.72cm]{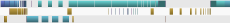}}
	\caption{Execution timeline for $P=8$ of GPU $p=0$ without (top) and with (bottom) overlapping communication with computation. The first row in each figure is the main computation stream, the second is the secondary communication stream, and the third is the low priority stream. Computational kernels are in blue, purple and cyan. Transfers to and from the CPU and GPU are in gold. Large gaps represent MPI communication.}
	\label{fig:streaming_distributed}
\end{figure*}

Figure \ref{fig:streaming_distributed} shows the effect of overlapping communication with computation on the timeline of the overall distributed matrix-vector multiplication. As expected, the gaps in the timeline due to communication are significantly smaller when it is overlapped with the computation.

%

\section{Algebraic Matrix Compression}
\label{sec:compression}

\Htwo{} algebraic matrix compression is an operation that takes as input an \Htwo{} matrix and produces another \Htwo{} matrix of lower rank that approximates the input to a desired target accuracy. Compression is akin to the truncation operation that is commonly used to approximate dense matrices or dense matrix blocks by low rank approximations.

Recompression is a core operation when working with \Htwo{} matrices. In particular, it is common in the discretization of integral equations to generate an initial \Htwo{} matrix from a kernel and an admissibility condition by approximating the kernel with Chebyshev polynomials for well separated clusters.  The ranks produced by this approximation are not optimal however, resulting in increased storage and increased arithmetic costs for operations such as the matrix-vector multiplication. A recompression step, purely algebraic, is then performed to generate an optimal basis in which the ranks are as small as possible. Another context where recompression is essential arises when adding matrices, performing low rank updates, or implementing BLAS3-like operations. When matrix blocks get added there is an increase in the apparent rank of the resulting blocks. The matrix would then need to be recompressed in order to maintain the linear asymptotic rate for storage and operations. The key task of recompression is to generate a new nested compressed basis in which the matrix blocks are expressed. This may be done by a pair of downsweep and upsweep operations that we describe next.

\subsection{Downsweep for Generating a New Basis}

Consider how a new basis for a block row $A_i^q$ at the finest level $q$ would be generated. $A$ here denotes only the low rank portions of the hierarchical matrix, since the dense blocks are not compressed.  $A_i^q$ consists of $b$ low rank blocks expressed at level $q$ as $U_i^q S_{ij}^qV_{j}^{q T}$ with $j=j_1 \cdots j_b$, and additional pieces representing the restriction of blocks from coarser levels to their ``i'' rows.
\begin{equation}
A_i^q = U_i^q \begin{bmatrix} \substack{\text{subblocks from} \\ \text{coarser levels}} \quad  S_{ij_1}^q V_{j_1}^{q T} \cdots S_{ij_b}^q V_{j_b}^{q T} \end{bmatrix} 
      = U_i^q B_i^{qT} 	
\end{equation}
A new more efficient basis can be generated by computing the SVD of $U_i^q B_i^{qT}$, and using the left singular vectors as the new basis. This would however require an expensive SVD of the $O(N)$-sized $B_i^q$. In order to avoid it, we first compute the QR decomposition of $B_i^q$ and then perform the SVD on the small $R$ factor. 
\begin{equation}
A_i^q =  U_i^q {B_i^q}^T = U_i^q (Q_i^q R_i^q)^T  = \underbrace{U_i^q {R_i^q}^T}_{\text{new basis}}  {Q_i^q}^T	= \overline{U}^q_i {Q_i^q}^T	
\end{equation}
The new basis for level $q$, $\overline{U}^q_i$ is effectively a reweighing of the columns of the previous basis $U_i^q$.

The task of computing $R_i^q$ of the QR decomposition of $B_i^q$ can be done efficiently by exploiting the nestedness of the bases.  Let us assume that the QR decomposition of $B_{i^+}^{q-1}$, the parent block $i^+$ at level $q-1$, is available as $Q_{i^+}^{q-1} R_{i^+}^{q-1}$. Then, 
\begin{equation}
\begin{split}
 A_i^q &=  \begin{bmatrix} \substack{\text{$i$-portion of} \\ U_{i^+}^{q-1} {B_{i^+}^{q-1}}^T}
          \quad U_i^q S_{ij_1}^q V_{j_1}^{q T} \cdots U_i^q S_{ij_b}^q V_{j_b}^{q T} \end{bmatrix} \\
	   & =  U_i^q \begin{bmatrix} E_i^q (Q_{i^+}^{q-1} R_{i^+}^{q-1})^T
          \quad S_{ij_1}^q V_{j_1}^{q T} \cdots S_{ij_b}^q V_{j_b}^{q T} \end{bmatrix} 
	   =  U_i^q B_i^{qT} 
\end{split}
\end{equation}
with $B_i^q$ conveniently expressible as:
\begin{equation}
	\label{eq:Btq}
B_i^q = \begin{bmatrix} Q_{i^+}^{q-1} \; R_{i^+}^{q-1} E_i^{q T} \\
	                    V_{j_1}^l S_{ij_1}^{q T} \\ \vdots \\ V_{j_b}^q S_{ij_b}^{q T}
        \end{bmatrix}
	  = \textbf{diag}(Q_{i^+}^{q-1}, V_{j_1}^q, \cdots, V_{j_b}^q)  
	    \begin{bmatrix} R_{i^+}^{q-1} E_i^{qT} \\ S_{ij_1}^{q T} \\ \vdots \\ S_{ij_b}^{q T}
	    \end{bmatrix}.
\end{equation}
Assuming the $V^q$ bases are orthogonal, the block diagonal matrix in Eq.~\ref{eq:Btq} is orthogonal and the QR  of $B_i^q$ simply reduces to the QR of the small stack at the end of Eq.~\ref{eq:Btq} which involves only $b+1$ blocks, each being a small $k \times k$ coupling/transfer matrix. Since this QR uses the $R^{q-1}$ matrix from level $q-1$, the overall computation starts from the root and goes down the tree computing all the $R^l_i$ matrices for all levels in a downsweep traversal. 

As with previous operations, all blocks at a given level can be processed in parallel, and importantly for the distributed setting, all the processes owning the subtrees below the C-level can proceed independently. Therefore, the computational pattern is identical to the distributed downsweep of the matrix-vector multiplication. Above the C-level, a single GPU is responsible for processing the top levels of the basis tree. The leaves of the top subtree, which hold the $R^l$ factors at level $l$, are then scattered to all GPUs and seed the roots of the individual subtrees, which continue the downseep independently all the way to the leaf level of the basis. The computational work at every level being, for every block row, a QR decomposition of the small stack at the end of Eq.~\ref{eq:Btq}. Batched QR operations \cite{boukaram18} are used within every GPU to achieve high performance.

\subsection{Upsweep for Truncating the New Basis}

Once the new reweighed basis is generated for all levels, it can then be truncated to the desired accuracy to generate the optimal basis $U^{\prime q}_i$. The coupling blocks are then compressed by projecting them on the new truncated basis. 

The truncation operation has to preserve the nestedness property. This can be accomplished by starting the truncation at the leaf level and sweeping up the basis tree. 
 Given the reweighed basis $\overline{U}$ ($\overline{U}_t$ at the leaves and $\overline{E}_t$ at nodes in the tree), we seek a basis $U'$ ($U'_t$ at leaves and $E'_t$ at nodes in the tree) that spans the same subspace as $U$ to the desired accuracy.
 
At the leaf level, this may be done through an SVD of the explicitly available basis
\[
\overline{U}_t = U \Sigma V^T \approx U'_t T_t 
\]
$U'_t$ is a subset of columns of the left singular vectors $U$. $T_t$ is a transformation matrix (which may be computed as $U'^T_t \overline{U}_t$) used to compute the new transfer matrix from a node to its parent as described next.

For non-leaves, the truncated basis is expressed in terms of new compressed transfer operators: 
\[
\overline{U}^{l-1}_t
=
\begin{bmatrix}  \overline{U}^l_{t_1} & \\ & \overline{U}^l_{t_2}  \end{bmatrix}
\begin{bmatrix}  \overline{E}^l_{t_1} \\ \overline{E}^l_{t_2}  \end{bmatrix}
=
\begin{bmatrix} U'^l_{t_1}  & \\ & U'^l_{t_2}  \end{bmatrix}
\begin{bmatrix}	T^l_{t_1} \overline{E}^l_{t_1} \\ T^l_{t_2}  \overline{E}^l_{t_2} \end{bmatrix}
\approx
\begin{bmatrix} U'^l_{t_1}  & \\ & U'^l_{t_2}  \end{bmatrix}
\begin{bmatrix}	E'^l_{t_1} \\ E'^l_{t_2} \end{bmatrix}
\begin{bmatrix}	T^{l-1}_t \end{bmatrix}
\]

$E'_{t_1}$ and $E'_{t_2}$ are the new transfer matrices for the children of a non-leaf node $t$, and $T_t$ is a transformation that will be used to compute the new transfer matrix from $t$ to its parent, in a similar fashion.
$\left[ \begin{smallmatrix} E'^l_{t_1} \\ E'^l_{t_2} \end{smallmatrix} \right]$ are computed by first computing an SVD of the matrix $\left[ \begin{smallmatrix}	T^l_{t_1} \overline{E}^l_{t_1} \\ T^l_{t_2}  \overline{E}^l_{t_2}  \end{smallmatrix} \right]$. The left singular vectors corresponding to singular values below the target compression threshold are truncated, and the remaining subset is partitioned to generate the new transfer matrices $E'_{t_1}$ and $E'_{t_2}$. $T_t$ is computed as $\sum_{t_i} E'^{lT}_{t_i} T^l_{t_i} E^l_{t_i}$.

The structure of the truncation algorithm is identical to that of the upsweep phase in the matrix-vector multiplication. All GPUs start the truncation operations concurrently, each on its subtree, with no interprocess communication required. The computational work at every level being, for every block row, an SVD involving the leaf basis at the leaf level, or the stacked transfer operators for non-leaf levels. Within a GPU, batched SVDs are used \cite{boukaram19a}. Once all GPUs reach the C-level, a gather operation communicates the new transfer operators from the roots of the subtrees to the leaves of the root tree that is stored on a single GPU. This bootstraps the last phase of the upweep which proceeds on the root GPU.

Two details remain to be discussed. First, in the downsweep phase, the algorithm relied on the basis $V$ being orthogonal. When this is not the case,  a pre-processing step is needed to orthogonalize the $\mathcal{V}$ basis tree. A basis is orthogonal if $V_j^{lT} V_j^l$ is the identity matrix for all levels $l$.  Orthogonalizing a basis involves performing QR on the finest level basis and then going up the tree to compute new transfer matrices that allow higher level nodes to satisfy the orthogonality condition. This is also done in an upsweep pass that is very similar to the one described above for truncation, but replacing the SVD operations by QR operations. The distributed implementation therefore also proceeds independently on all GPUs up until the C-level. A gather operation is then performed to bootstrap the orthogonalization of the top levels of the basis tree which reside on a single GPU.

Finally, once a new compressed basis $U'$ is computed, it remains to approximate the coupling blocks in it. Since $U'$ is an orthogonal basis, the best approximation of a block is obtained by an orthogonal projection. Therefore, we can obtain the approximation $A'$ by projecting every low rank block $A_{ts} = U_{t} S_{ts} V_{s}^T$ on the new basis to obtain:
\[ 
A'_{ts} = U'_{t} U'^T_{t} A_{ts} V'_{s} V'^T_{s}  = 
U'_{t} U'^T_{t} ( U_{t} S_{ts} V_{s}^T) V'_{s} V'^T_{s} = U'_{t} (P_{Ut} S_{ts} P_{Vs}^T) V'^T_{s} = U'_{t} S'_{ts} V'^T_{s} 
\]
The new $S'_{ts}$ coupling blocks are computed via batched matrix multiplication operations. In this step, there is parallelism across all GPUs and across all levels. In addition, The GPUs are particularly efficient in GEMM operations, and in practice this projection step consumes much less time that the other operations, particularly those involving batched SVDs. 

\section{Performance Results}
\label{sec:results}

In this section, we report performance results for the distributed-memory matrix-vector multiplication and algebraic compression operations, as well as performance on a complete application involving the setup and solution of an integral equation. 

\subsection{Experimental Setup}

The H2Opus library as well as the code to generate and run all examples below is open source and is available at {\tt https://github.com/ecrc/h2opus}. Batched kernels for GEMM operations are executed using MAGMA \cite{baboulin08,magma}, while SVD and QR batches are performed using the algorithms shipped with the open-source library KBLAS, available at {\tt https://github.com/ecrc/kblas-gpu}.
The marshaling GPU kernel and various other utility routines use Thrust \cite{thrust19}.

All tests are conducted on Summit, the IBM Power System AC922 supercomputer installed at Oak
Ridge National Laboratory.
Individual nodes on Summit have 6 NVIDIA V100-SXM2 GPUs with 16GB of HBM2 memory each, but we only used 4 GPUs per node (2 per socket) in our runs.
Summit has a fast host-to-device bandwidth which can deliver up to 50 GB/s over PCIe 4.0; in our application, we were able to use a significant portion of it, 40 GB/s, as measured by {\tt nvprof}. It also has a fat-tree topology network for internode communication that delivers 200Gb/s of bandwidth. 
To assess the efficiency attained by our algorithms, we measure the performance of the single GPU batched GEMM implementation from MAGMA, with batch elements of size $64\times64$. 
All computations are done in double-precision and every point in every plot has been generated as the average of 10 runs after discarding the fastest and slowest timings.

To test the performance and scalability of the matrix-vector multiplication and matrix compression implementations, we performed numerical experiments on two sample matrix sets with different structural characteristics.
The first matrix set comes from a spatial statistics application using a point set placed on a 2D grid of side length $a$, and an exponential kernel with a correlation length of $0.1a$.
The hierarchical matrix representation of this covariance matrix uses $m=64$ as the finest block size and size of the dense leaves. A geometric admissibility condition $\eta ||C_t - C_s|| \ge (D_t + D_s)/2$ is used, where $C$ and $D$ refer to the center and diagonal size of a bounding box of the corresponding point set. We use a value of $\eta = 0.9$ and set a rank $k=64$ in the low rank blocks, resulting in an approximation with relative accuracy better than $10^{-7}$ for all problem sizes. This accuracy is computed by sampling $10\%$ of the rows and computing $||Ax - A_{\mathcal{H}^2} x|| / ||Ax||$ with randomly generated vectors with entries from a uniform distribution. The resulting sparsity constant of the matrix, which is a proxy for how finely refined the matrix is in its off-diagonal portions, is $C_{sp} = 17$.
At the largest size, the matrix has 23 levels, with the top 10 levels on a single master GPU, and the bottom 13 levels on separate 1024 GPUs for a matrix size of 536M. 

The second matrix set comes from a 3D Gaussian process and is intended to show the effect of memory pressure---due to a finer refinement in the off-diagonal blocks---on scalability. The matrices are constructed using a set of points on a 3D grid of side length $a$ and uses an exponential kernel with correlation length $0.2a$ \cite{ambikasaran16}, a similar admissibility condition to the previous case, and a rank $k=64$ for the low rank blocks. The resulting relative accuracy is now $10^{-3}$ for all problem sizes considered in this section, and the resulting matrix tree has many more leaves than in the 2D case with a larger sparsity constant $C_{sp} = 30$.

\subsection{Matrix-Vector Multiplication}

%
%

\subsubsection{Weak Scalability}
\begin{figure}[ht]
\includegraphics[width=0.32\textwidth]{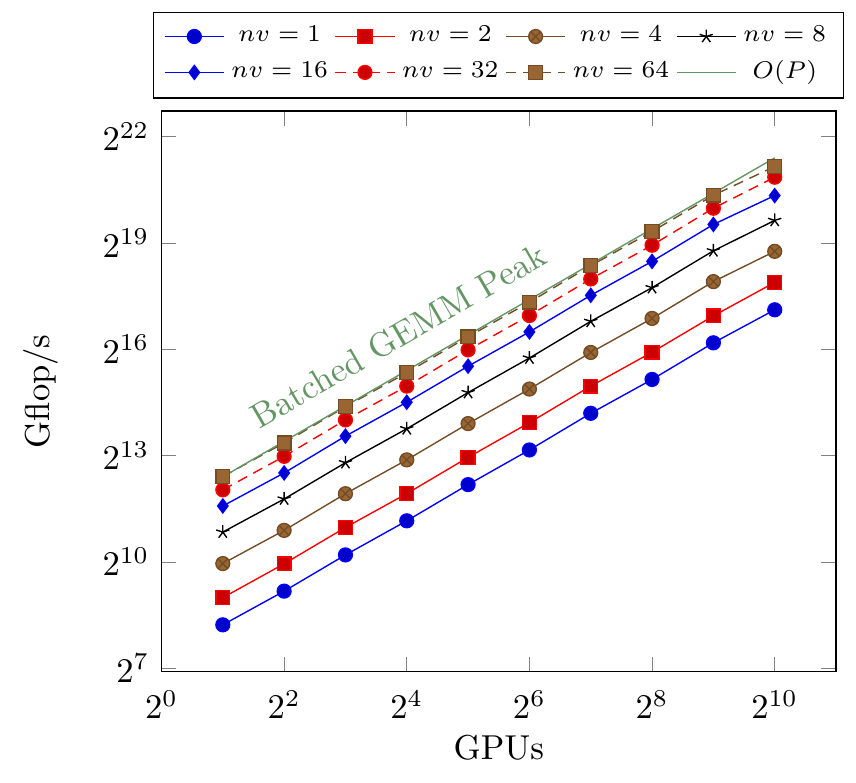}
\includegraphics[width=0.32\textwidth]{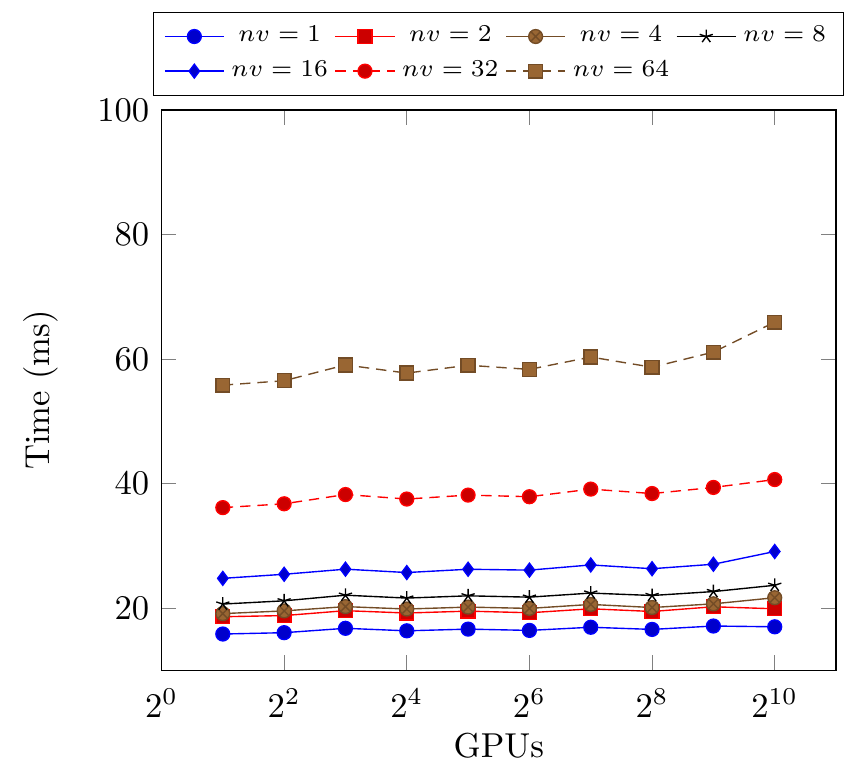}
\includegraphics[width=0.32\textwidth]{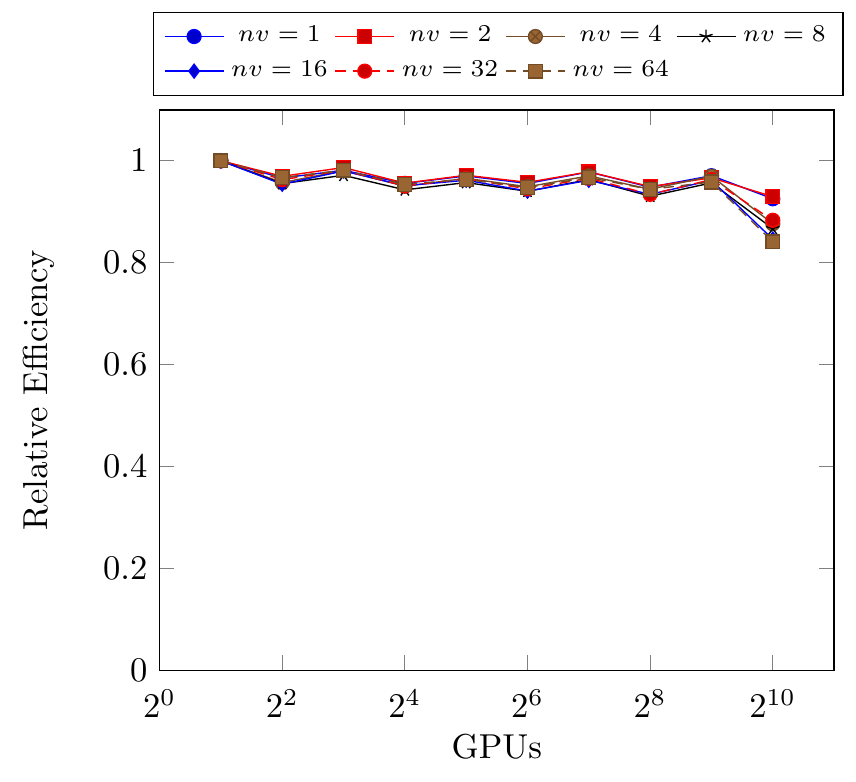}

\includegraphics[width=0.32\textwidth]{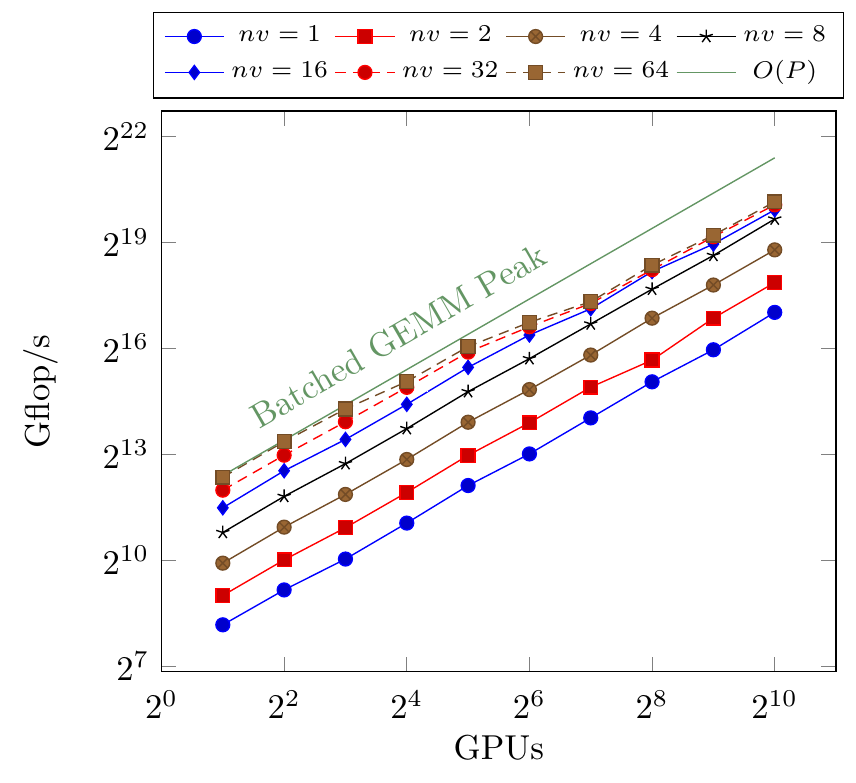}
\includegraphics[width=0.32\textwidth]{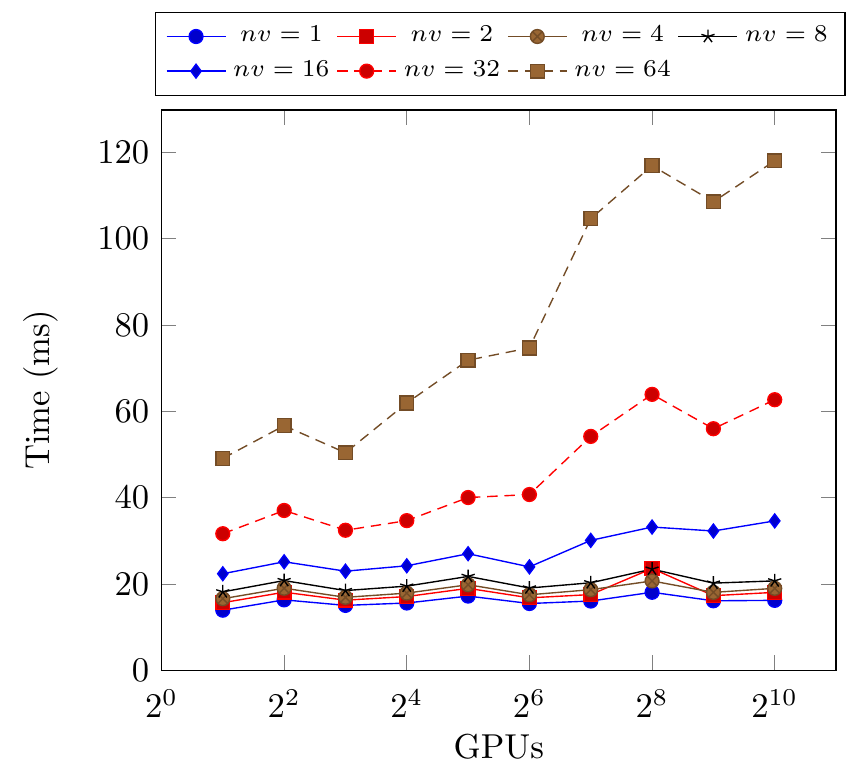}
\includegraphics[width=0.32\textwidth]{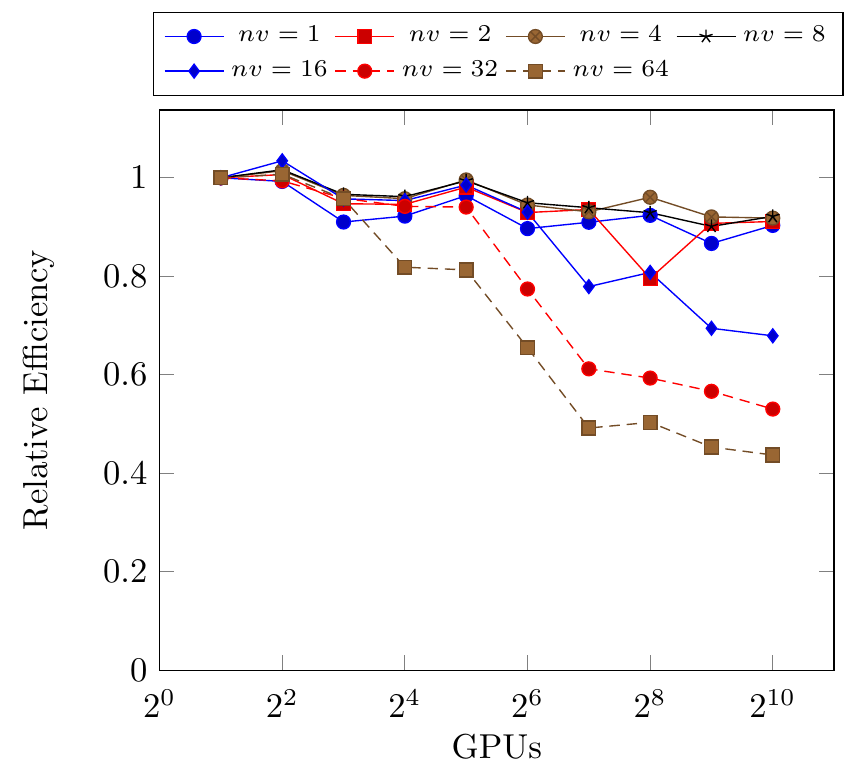}
\caption{Scalability and efficiency of HGEMV on Summit, using matrices constructed from 2D (top) and 3D (bottom) exponential kernels.}
\label{fig:summit_results}
\end{figure}

We first report on the weak scalability of the implementation,
using a number of vectors $nv$ ranging from 1 to 64.
Both 2D and 3D test sets were scaled using a local matrix size of $\pp{N}{p} = 2^{19}$ per GPU. Results are summarized in Figure \ref{fig:summit_results}, with the top and bottom rows showing the results from the 2D and 3D test sets, respectively. We observed no significant variability in the timings, with the highs and lows within 1--3\% of reported average. There is a slight jitter in the plots as the problems are scaled up, due to small changes in the structure of the matrix tree affecting the amount of actual computations performed, which do not grow exactly at the same rate as the problem size. The relative efficiency was computed as $(G_P/G_{P_0}) / (P/P_0)$, the ratio of the relative flops performed and the scaling factor relative to a base case of $P_0 = 2$ GPUs. 

For the 2D tests, the scalability is near ideal up to 512 GPUs across all multiple vector sizes, $nv = 1 \dots 64$. For the single vector case, a bandwidth-limited operation, the throughput is about 150 Gflop/s per GPU. With $nv = 64$, the additional arithmetic intensity pushes the computation into a compute-bound regime and achieves 2.6 Tflop/s per GPU, more than 95\% of the sustained peak of the batched GEMM operation measured at 2.7 Tflop/s. Even with the larger data volume being communicated, corresponding to various parts of 64 vectors, the communication is essentially hidden by the local computations. Only with 1024 GPUs, the plots show a slight degradation in performance and a deviation from ideal scalability, particularly for the $nv = 64$ case, with performance dropping to 2.3 Tflop/s per GPU. The primary reason for this is that the root tree on the master GPU is now 10 levels deep and is of size $2^{16}$. The computations on this sizeable top tree become a bottleneck. In order to scale up to this number of GPUs or larger, we will need to coarsen to a single top GPU more smoothly than the $P$-to-$1$ ratio used in the current implementation. For example, the trees above a first C-level can be distributed on multiple GPUs, and back to a single GPU after a second C-level, so that the top level work is small enough to be overlapped with other parts of the computation.

The results of the 3D problem test set display a similar behavior, but the communication overhead appears earlier. For the single vector case, results scale reasonably well with problem size and the relative efficiency is about 90\% with 1024 GPUs, with a performance of about 130 Gflop/s per GPU. As the number of vectors $nv$ increases however, the communication needed for transferring all the relevant portions of the $\pp{\widehat{x}}{p}$ vectors to the GPUs that need those data becomes substantial. This is in addition to the larger root tree difficulty mentioned above. The compute part of the operation grows sub-linearly with problem size because of the favorable increase in arithmetic intensity; however, the communication volume grows linearly and can no longer be hidden by the now relatively faster compute phases. While a performance of 2.6 Tflop/s per GPU is reached for 2 GPUs, the performance at 1024 GPUs reaches only 1.1 Tflop/s, an efficiency less than 45\%. 

\subsubsection{Strong Scalability} 
\begin{figure}
\includegraphics[width=.49\columnwidth]{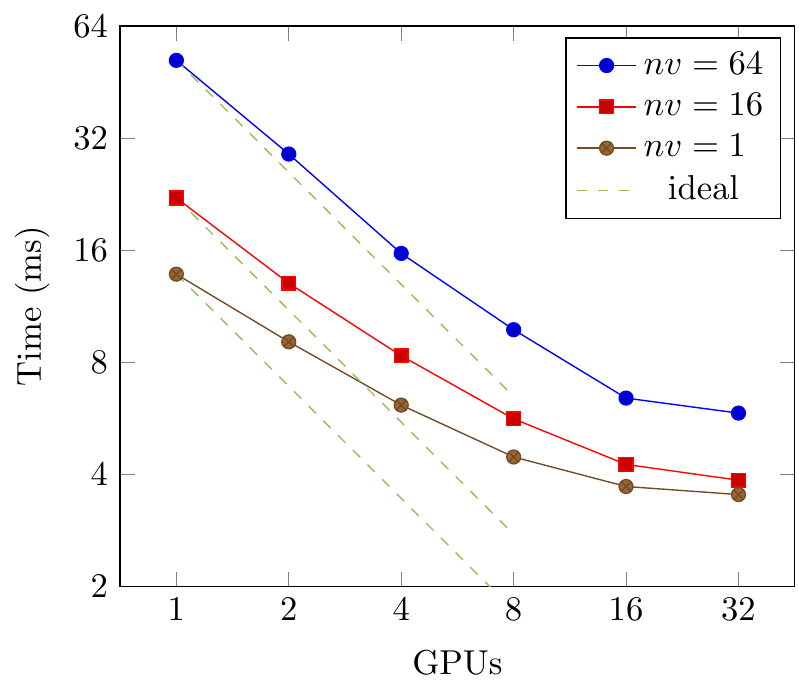}
\includegraphics[width=.49\columnwidth]{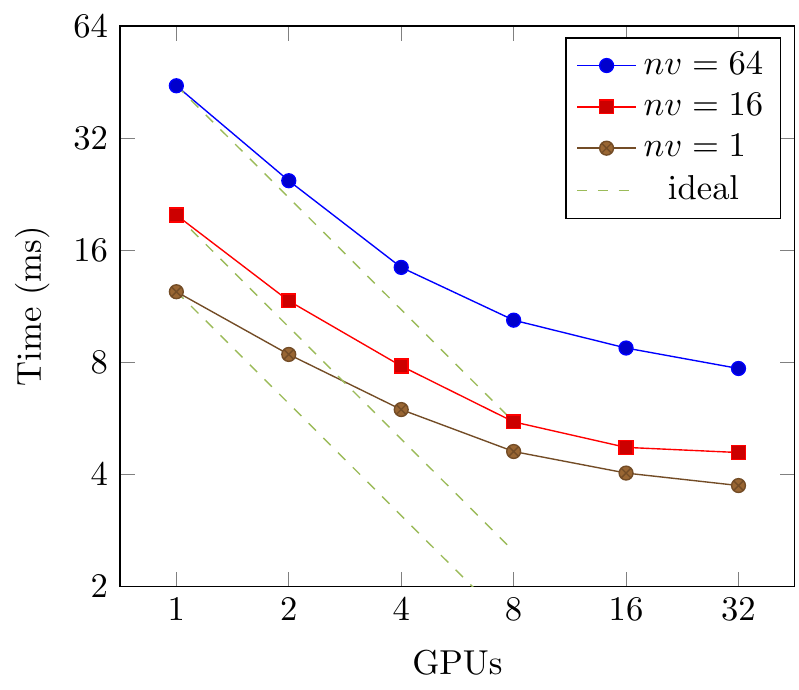}
\caption{ Strong scalability results for 2D (left) and 3D (right) test data.} \vspace*{-8pt} 
\label{fig:ss}
\end{figure}  
We report, in Figure \ref{fig:ss}, strong scalability results of the matrix-vector multiplication for the 2D and 3D test data described above, for different numbers of test vectors. In all cases, the problem size is $N = 2^{19}$ to fit on a single GPU of Summit and is executed on an increasingly larger number of GPUs. As expected, the more arithmetically intensive multiplication involving a larger set of vectors scales better than the single vector operation. In all cases however, the limits of strong scalability is reached around 32 GPUs. This is not unexpected, since at this scale the local problem size is now only $\pp{N}{p} = 2^{14}$. There is very little local work to do to hide communication, and the whole computation takes only few milliseconds.

\subsection{Matrix Compression}

To test the performance and scalability of the algebraic compression routines, we used the same sets of matrices described earlier. For the 2D tests, we start with the matrix defined as above, with point clusters of size $m = 64$, and admissibility parameter $\eta = 0.9$. Its low rank blocks are initially constructed using a Chebyshev polynomial approximation of the kernel in the bounding boxes of the point clusters. A $6 \times 6$ grid is used, resulting in all low rank blocks having a uniform rank $k = 36$. Compression seeks to reduce these ranks to maintain an accuracy of $\tau = 10^{-3}$. The 2D test set was scaled using a local matrix size $\pp{N}{p} = 2^{20}$, reaching a matrix size of 67M on 64 GPUs. 

For the 3D tests, a point cluster size $m = 64$ and an admissibility parameter $\eta = 0.95$ are used in the matrix construction. The low rank blocks are initially computed using a tri-cubic Chebyshev polynomial approximation of the kernel in the bounding boxes of the point clusters, resulting in all low rank blocks having a uniform rank $k = 64$.  Compression seeks to reduce these ranks to maintain an accuracy of $\tau = 10^{-3}$. The test test was scaled using a local matrix size $\pp{N}{p} = 2^{18}$, reaching a matrix size of 16M on 64 GPUs.

\begin{figure}[ht]
	\includegraphics[width=.325\textwidth]{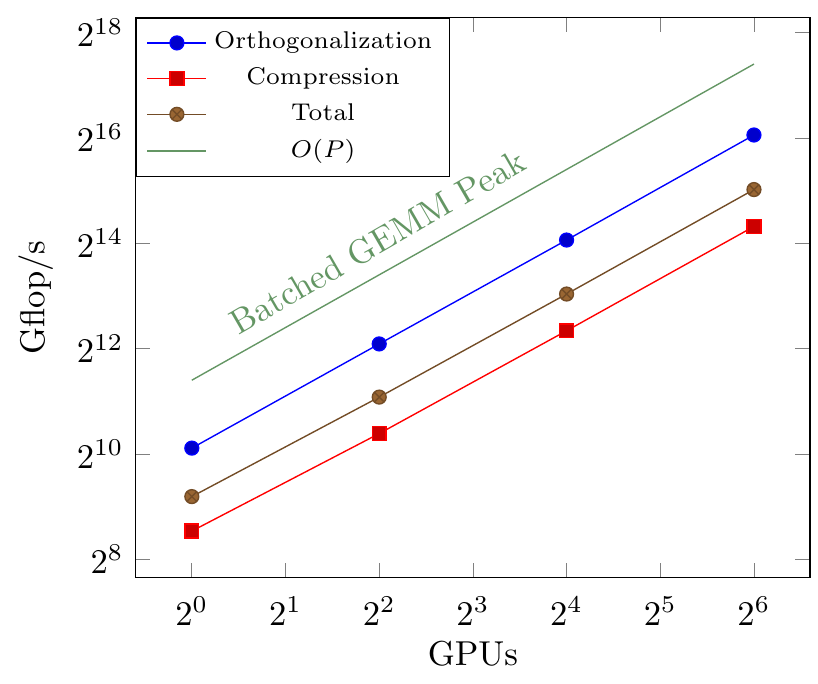}
  \includegraphics[width=.325\textwidth]{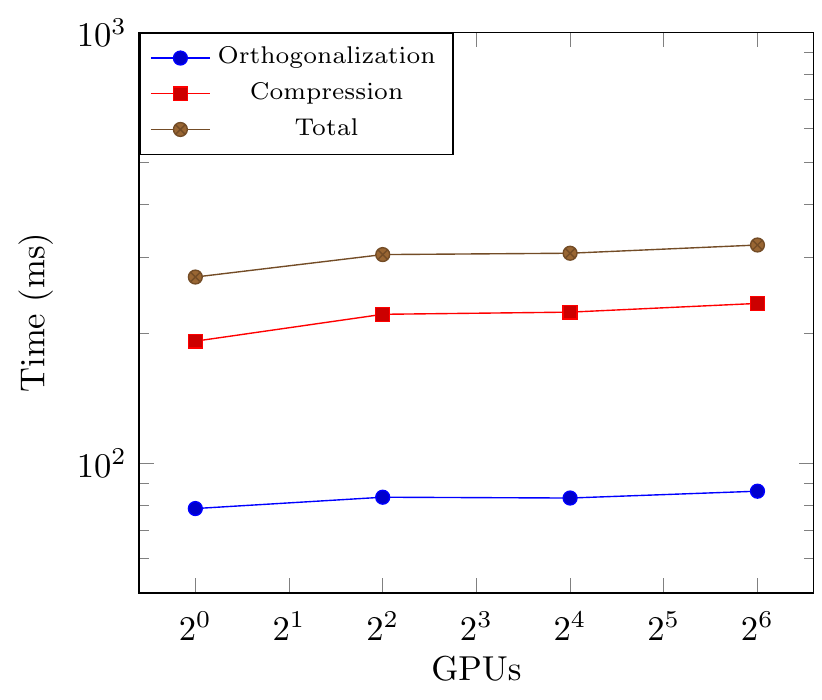}
  \includegraphics[width=.325\textwidth]{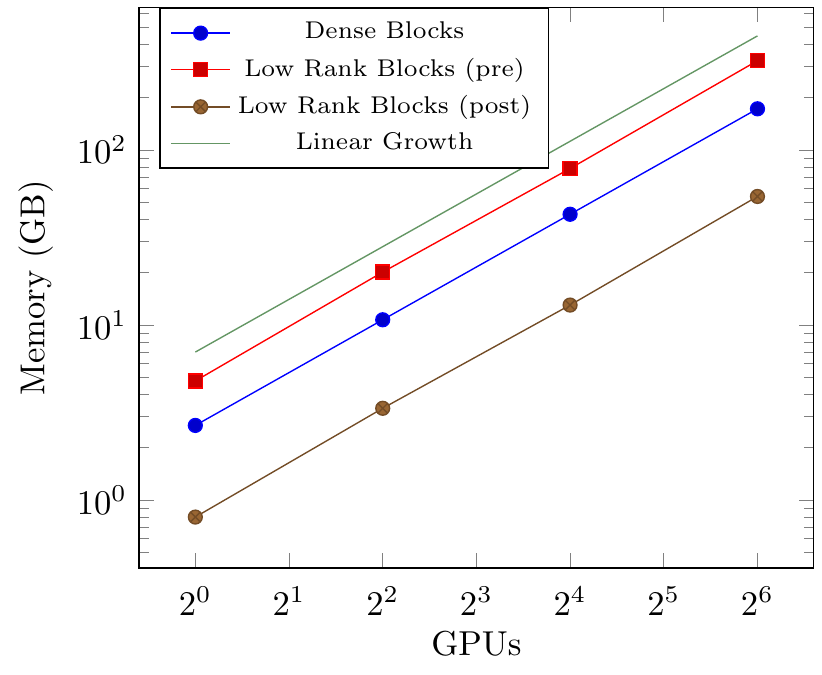}
	\includegraphics[width=.325\textwidth]{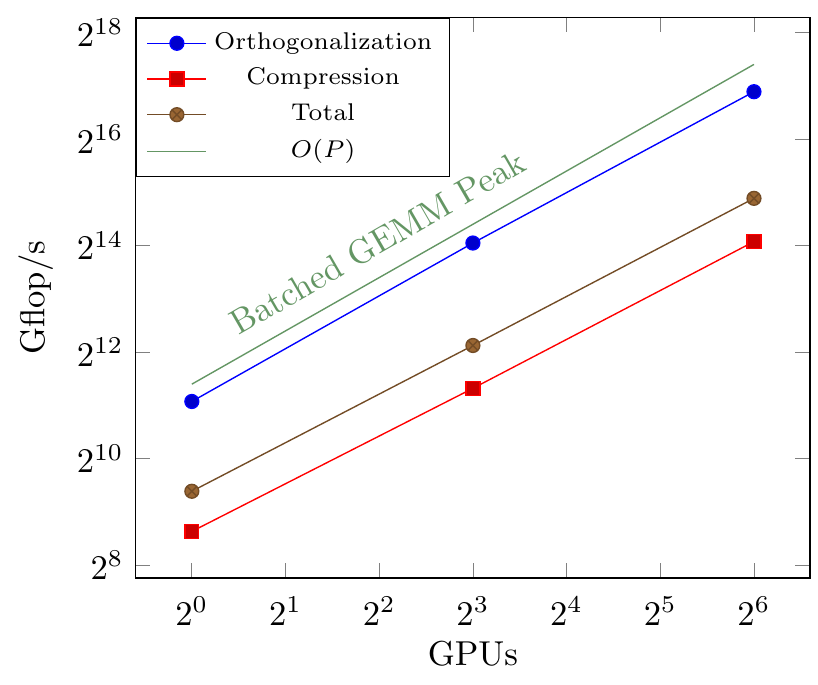}
  \includegraphics[width=.325\textwidth]{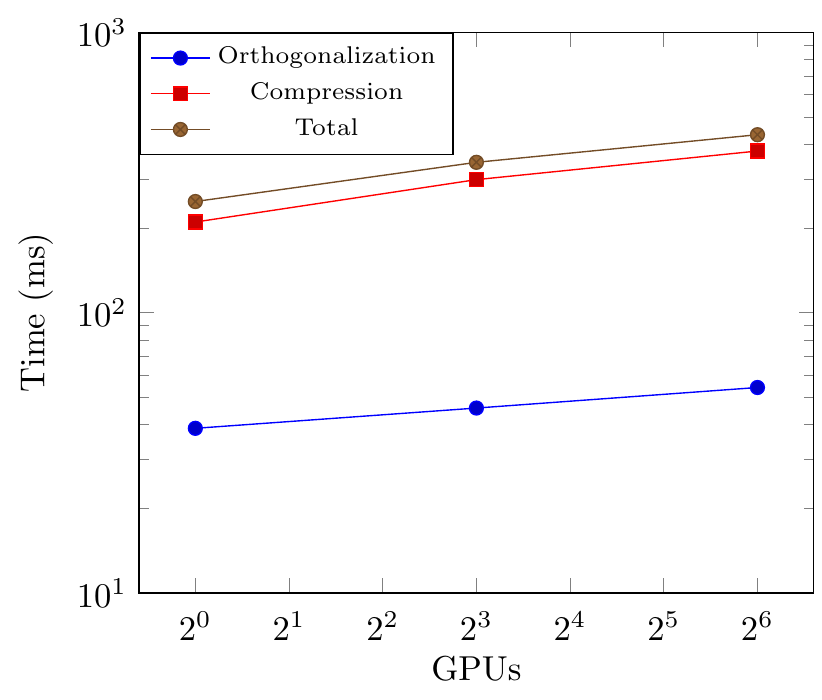}
  \includegraphics[width=.325\textwidth]{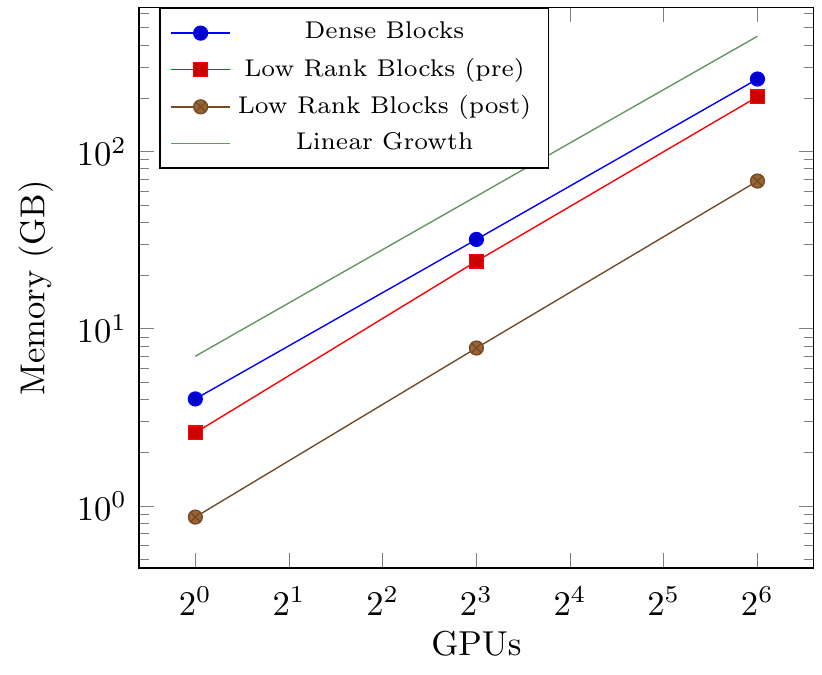}
  \caption{Scalability of algebraic compression for 2D (top) and 3D (bottom) test sets.}
  \label{fig:compression} 
\end{figure}

\subsubsection{Weak Scalability} 
Figure \ref{fig:compression} shows the weak scalability and effectiveness of the compression operation. Compression starts with an initial phase to orthogonalize the (non-orthogonal) bases constructed by Chebyshev interpolation without modifying their ranks. We time and report the orthogonalization phase separately. This is followed by a pair of downsweep/upsweep passes to construct a new reweighed basis and truncate it, finalized with a projection of the low rank blocks on the new bases. These steps are reported together under the compression phase label.  The orthogonalization phase involves fewer floating point operations and as a result it executes faster than the compression phase. In 2D, the scaling is near ideal for up to 64 GPUs, with a slight bump when moving beyond a single GPU, because of inter-GPU communication. However for $P > 1$, the execution time is essentially flat indicating that all communication has been hidden by the computational steps. In 3D, we note a slight deviation from ideal scalability when $P = 64$, because the communication volume is larger than the 2D case and it can no longer be totally hidden by local computations. The orthogonalization phase executes 
at around 2.1 Gflop/s/GPU (near the practically-achievable peak of 2.7 Gflop/s per GPU, as measured by the batched GEMM operation). The compression phase, featuring QR on stacks of $C_{sp}$ blocks and SVD kernels, is not able to reach that level of performance, but this limitation comes purely form the single GPU performance of the corresponding batched routines for these operations, and would be directly improved when more performant batch kernels are implemented. Nonetheless, we should note that even with our current batch kernels, the 67M and 16M matrices in 2D and 3D respectively, are compressed in just a fraction of a second, as reported in the central column of Figure \ref{fig:compression}.

In the rightmost column of Figure \ref{fig:compression}, we report on the effectiveness of the compression operation in reducing the memory footprint of the low rank blocks. In the 2D case, there is a factor of $6\times$ reduction between pre-compression low rank memory (with all blocks having a uniform rank $k = 36$) and post-compression to an accuracy of $\tau = 10^{-3}$. In the 3D case, the compression is a factor of $3\times$ for the low rank data, primarily because the starting matrix was generated with a relatively small footprint (all blocks have rank 64, generated from tri-cubic polynomials) and not much reduction is possible to maintain an accuracy of $\tau = 10^{-3}$. In all cases however, we note the $O(N)$ ideal growth in memory. 

\subsubsection{Strong Scalability} 
\begin{figure}[ht]
  \begin{center}
	\includegraphics[width=.45\textwidth]{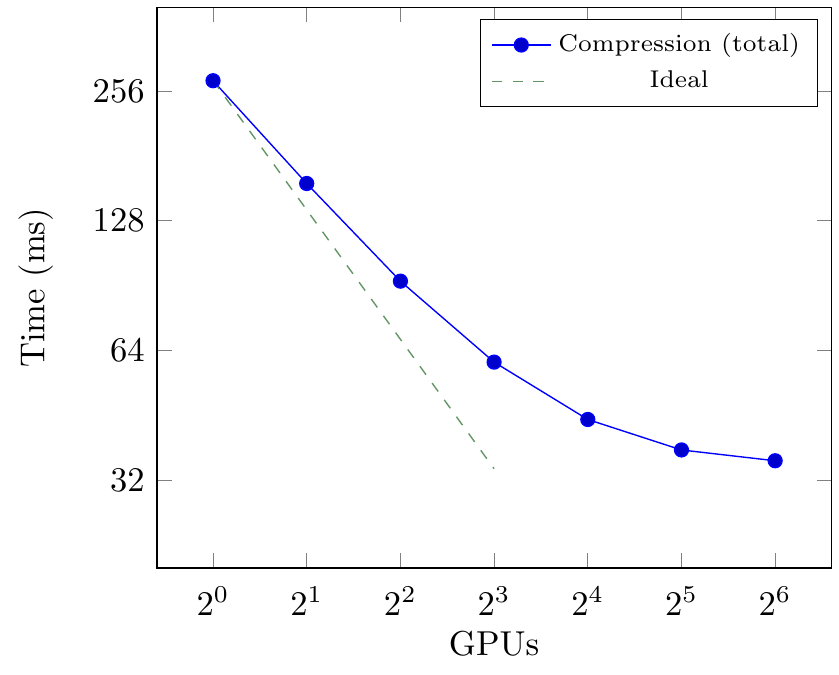}
  \includegraphics[width=.45\textwidth]{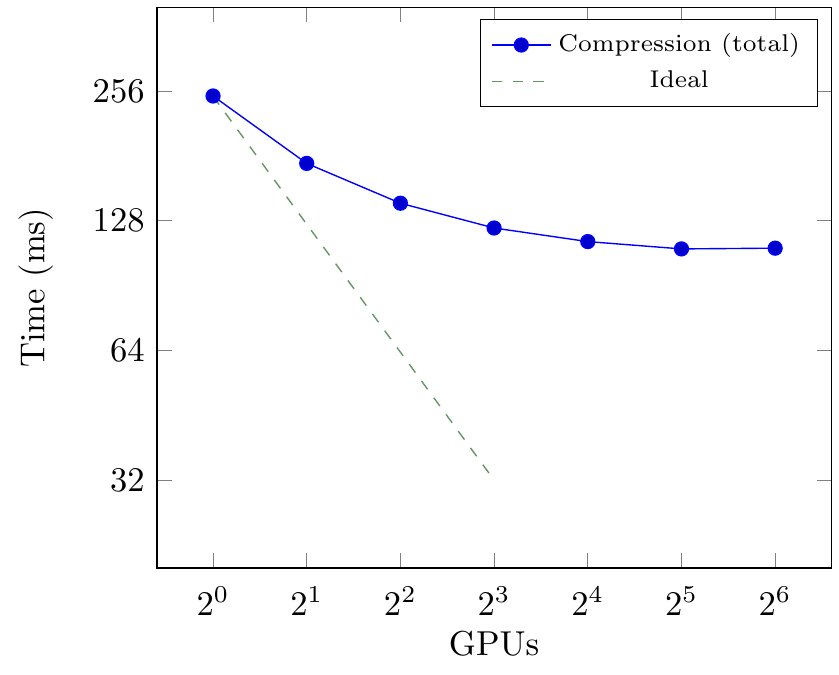}    
  \end{center}
  \caption{Strong scalability of algebraic compression for 2D (left) and 3D (right) test sets.}
	\label{fig:compression_strong_scaling} 
\end{figure}

Strong scalability results are shown in Figure \ref{fig:compression_strong_scaling}. Deviation from ideal scalability becomes noticeable as soon as the local problem size is small enough where there are not enough computations to perform locally and communication time dominates. For the 2D tests, the problem size is $2^{20}$. With $P = 8$ GPUs, the local problem size is only $\pp{N}{p} = 2^{17}$ and results in an efficiency reduction to near $50\%$. On 32 GPUs, the local problem size is $\pp{N}{p} = 2^{15}$ and the limit of strong scalability is essentially reached: there is very little local work to do and the whole operation takes a few ms, spent mostly in communications. For the 3D tests, the problem size is $2^{18}$. On $P = 4$ GPUs, the local problem size is already only $\pp{N}{p} = 2^{16}$, and the resulting efficiency drops below $50\%$. With 16 GPUs, the strong scalability limit is essentially reached as the problem size is now only $2^{14}$, with very little local work available for each GPU.

\subsection{Integral Fractional Diffusion Equation}

We consider the performance of a Krylov solver for the solution of the integral equation $\mathcal{L}[u(\x)] = b(x)$ where $\mathcal{L}$ is the fractional diffusion operator defined as:
\begin{equation}
   \mathcal{L}[u(\x)] = 
       -2 \int_{\Omega \cup \Omega_0} \frac{u(\y) - u(\x)}{|\y - \x|^{n + \beta(x) + \beta(y)}} a(\x, \y) d\y
  \label{eq:integral_nd}
\end{equation}

The spatially varying diffusivity $a(\x, \y)$ is defined as the geometric mean of the usual diffusion coefficient, $a(\x, \y) = \kappa(\x)^{1/2} \kappa(\y)^{1/2}$, $\beta$  is the fractional order ($0.5 < \beta < 1$) and we assume it to be constant in this experiment, $\Omega \in R^n$ is the region in which the solution $u$ is sought, and $\Omega_0 \in R^n$ is a surrounding region in which volume constraints, somewhat analogous to the Dirichlet boundary conditions in the classical diffusion equations, are imposed on the solution, $u(\x) = 0$ for $\x \in \Omega_0$.  We solve the problem with $\Omega = [-1, 1]^2$, $\Omega_0 = [-3, 3]^2 \setminus [-1, 1]^2$, and constant fractional order $\beta = 0.75$. We use a diffusivity field of the form
\begin{equation}
  \kappa(\x) = 1 + f(\x_1; 0, 1.5) ~f(\x_2; 0, 2.0)
\end{equation}
with the bump function $f$ defined as: 
\begin{equation}
  f(x; c, \ell) = 
  \begin{cases}
    \mathrm{exp}\big(\! - \! \frac{1}{1-r^2}\big), \, r = \frac{x - c}{\ell /2}, & |r| < 1 \\
    0, & |r| \ge 1
  \end{cases}
\end{equation}
and a right hand side $b(x) = 1$ in $\Omega$. 

The singularity of the kernel in (\ref{eq:integral_nd}) requires that the discretization of the integral be done carefully to allow standard quadrature rules to attain their theoretical convergence rate. To this end, we rewrite the integral as: 
\begin{align}
&  \int_{\Omega \cup \Omega_0} 
\left[ \frac{-2\left[u(\y) - u(\x)\right] \, a(\x,\y)}{|\y-\x|^{n+2\beta}} + 
\frac{p_{\x}(\y)}{|\y-\x|^{n+2\beta}} \right] d\y 
\, - 
\int_{\Omega \cup \Omega_0} 
\frac{p_{\x}(\y)}  {|\y-\x|^{n+2\beta}} \, d\y
\label{eq:regularized}
\end{align}
where $p_{\x}(\y)$ is a function with local support around $\x$ chosen to remove the singularity of the integrand and allow
a trapezoidal rule to achieve its quadratic convergence. The first integral in (\ref{eq:regularized}) can be then discretized on a regular grid while the second integral involves terms that can be integrated analytically.  The details are derived in \cite{alzahrani21}, which generalizes the treatment of \cite{minden20} to the variable coefficient case. 
This correction has the property that it allows the hierarchical matrix treatment of the kernel to remain as is, similarly to the FMM-compatible treatment of integral equations on curves \cite{hao14,wu21}  and other locally-corrected quadrature schemes \cite{greengard21}.

In 2D, using a regular grid of $N$ points with spacing $h$ in $\Omega$, the final discretization results in a system of the form:
\begin{equation}
  h^2(D + K + C) u = b
\end{equation}
where $D$ is an $N \times N$ diagonal matrix with entries
\begin{equation}
D_{ii}=  \sum_{\substack{j \ne i \\ \y_j \in \Omega \cup \Omega_0}}  \frac{2 a(\x_i,\y_j)}{|\y_j-\x_i|^{2+2\beta}},
\end{equation}
$K$ is an $N \times N$ formally-dense matrix with zero diagonals and entries
\begin{equation}
K_{ij} = -\frac{2 a(\x_i,\y_j)}{|\y_j-\x_i|^{2+2\beta}}, \ \text{for } i \ne j,
\end{equation}
while $C$ is a sparse matrix resulting from the regularization of the integral. $C$ is essentially the discretization of an inhomogeneous, but non-fractional, diffusion operator, and we use it to construct a preconditioner for (\ref{eq:regularized}). Its footprint on a regular grid is the same as a 5-point Laplacian discretization, for an $O(h^2)$ accuracy. (Note however that when the solution does not have sufficient regularity, only first-order convergence in solution error can be expected).

The matrix $K$ is constructed and compressed as an \Htwo{} matrix using the H2Opus facilities described above. A $k$-d tree partitions the $N$ grid points into clusters that are distributed to the different GPUs. An admissibility condition defines the structure of the matrix. Dense blocks are generated by evaluating the kernel directly. The low rank blocks are first approximated using polynomial Chebyshev approximations of the kernel on bounding boxes of the point clusters, and they are then algebraically compressed to generate the matrix that is used during the solution phase.   

The diagonal matrix $D$ is constructed by noting that its entries can be recast as the product $\hat{K} \mathbf{1}$, where $\hat{K}$ is a matrix similar to $K$, but defined on the larger region $\Omega \cup \Omega_0$, discretized with the same $h$, and $\mathbf{1}$ is a vector of ones. We therefore use the distributed-memory facilities of H2Opus to generate $\hat{K}$, perform the matrix-vector multiplication, to compute the diagonal entries of $D$. $\hat{K}$ is then discarded. 

Finally, the sparse matrix $C$ is distributed among the processors using the same blockwise row partitioning of the matrix $K$, and its entries computed explicitly. 

The Krylov solver uses the facilities of the PETSc library. The operator to  be inverted is symmetric positive definite, and we thus employ a preconditioned conjugate gradient method as solver. For the preconditioning step, we consider a smoothed aggregation algebraic multigrid method constructed on the matrix $C$, using a diagonally preconditioned Chebyshev method as a smoother.
The setup of the preconditioner runs partly on the  CPU and  partly on the GPU, while the Krylov solver, including  the  preconditioner application, runs entirely on the GPU \cite{mills2020toward,zhang2021petscsf}.

\begin{figure}[ht]
	\includegraphics[width=.45\textwidth]{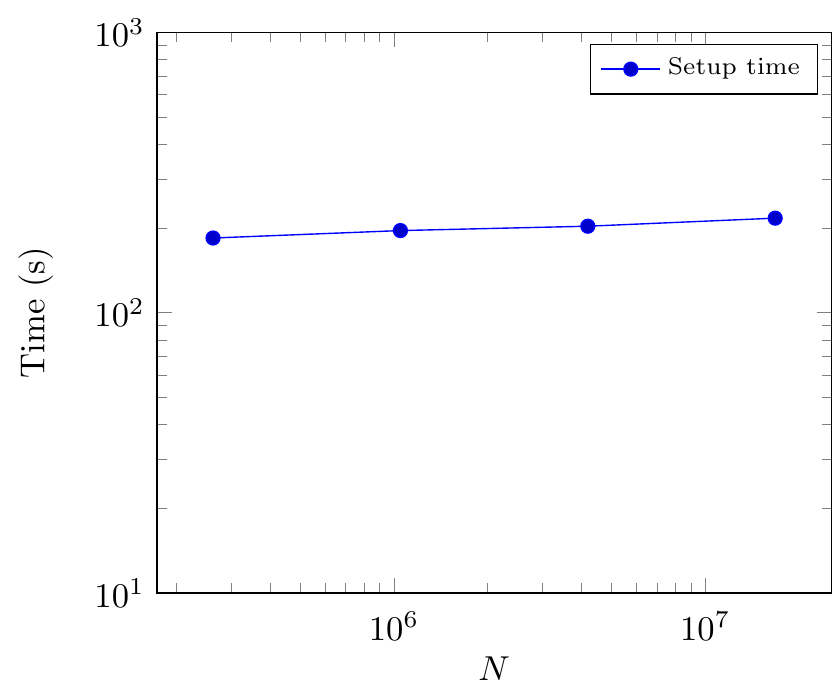}
	\label{fig:ws_setup}
  \includegraphics[width=.45\textwidth]{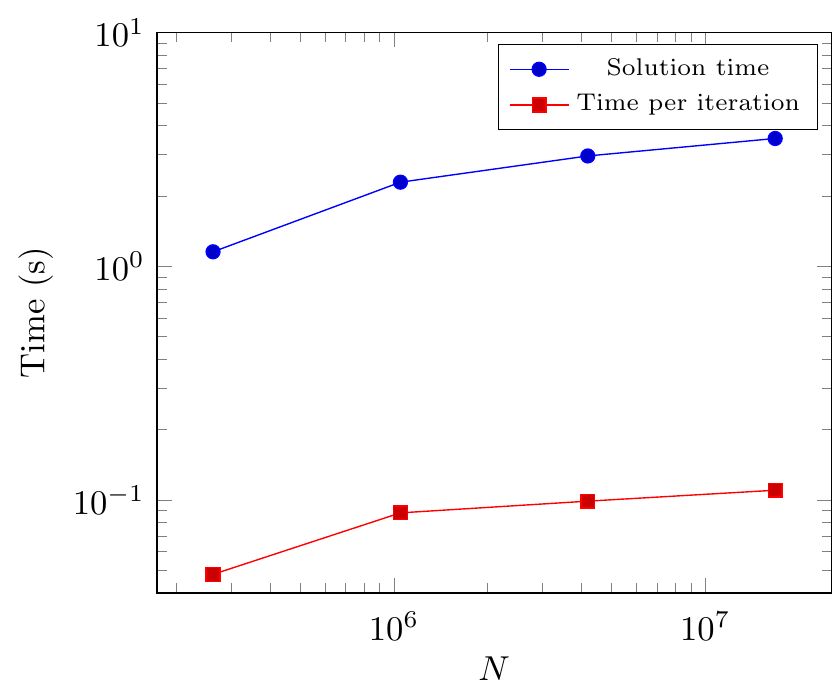}
	\label{fig:ws_solve}  
  \caption{Performance of integral fractional diffusion solver in 2D}
  \label{fig:solver}
\end{figure}

The weak scalability of the solver is shown in Figure \ref{fig:solver} on  grids of size $512 \times 512$, $1024 \times 1024$, $2048 \times 2048$, and $4096 \times 4096$ using 1, 4, 16, and 64 GPUs respectively. 
Setup times are shown in the left panel as a function  of the size of the problem $N$, and they include all the necessary steps to assemble the operator to be inverted together with the setup of the preconditioner. The only setup computation not included in the timing is the construction of the $k$-d tree for the matrix structure generation. This is an $O(N \log^2 N)$ computation that is currently performed sequentially on the CPU. In the right panel, we report the total time for the Krylov solver, and the time per iteration as the problem size $N$ is scaled up.

Perfect weak scaling can be observed for the operators' setup; the increase in total solving time can be attributed to a slight increase in  the number of Krylov iterations being 24, 26, 30 and 32 using 1, 4, 16, and 64 GPUs respectively. Since the domain  $\Omega$ is kept fixed in the above tests while the discretization step is reduced, these results confirm the algorithmic scalability of the solver as well as of its parallel implementation. We are not aware of other software libraries that can readily handle variable coefficient fractional diffusion problems at this scale. 

\section{Conclusions}
\label{sec:conclusions}

We presented distributed-memory, GPU-accelerated, algorithms and implementations for two key operations on \Htwo{} matrices. The first is the matrix-vector (and multi-vector) multiplication and the second is the algebraic matrix re-compression to reduce the memory footprint consistent with a desired accuracy.  The algorithms are supported by distributed data structures for representing, accessing, and operating on hierarchical matrices with nested bases, and the inter-process MPI communication is optimized to hide much of the data transfer cost with local compute phases of the algorithms.  Both algorithms exhibit $O(N)$ behavior and near-ideal weak-scalability for a large number of GPUs; high performance on individual GPUs is achieved through the use of batched dense linear algebra kernels. 
The algorithms are incorporated in the open source library H2Opus and have interfaces to the PETSc library.

The performance and usefulness of the library is demonstrated through the solution of a variable diffusivity integral fractional diffusion problem in 2D. Algorithmic and implementation scalability are demonstrated on grids of size up to $4096\times 4096$ on 64 GPUs, using the facilities of the PETSc library for the Krylov solver and the construction of an algebraic multigrid preconditioner.

Future work is planned on three fronts. The first is the development of distributed-memory multi-GPU versions of additional \Htwo{} matrix algorithms, including their construction from randomized sampling. This will open the door for matrix-matrix multiplication and inversion via Newton-Schulz iterations. The second is the development of interfaces to BEM++ \cite{smigag15} and related high-productivity software libraries for integral equations that handle discretization tasks. H2Opus can provide highly effective core solvers for such libraries and will allow their use for large scale problems. The last front entails portability, and we plan to extend the H2Opus library to run on AMD GPUs.

\newpage


\end{document}